\documentclass{ws-p8-50x6-00}

\begin{document}

\title{Neutrino Oscillations in a "Predictive" SUSY GUT\footnote{This talk is based on work done in collaboration with T. Blazek and K. Tobe.}\footnote{Ohio State University preprint OHSTPY-HEP-T-99-014}}

\author{Stuart Raby}

\address{Department of Physics, The Ohio State University, Columbus, OH 43210,
USA\\E-mail: raby@mps.ohio-state.edu}

%%%%%%%%%%%%%%%%%%%%%%%%%%%%%%%%%%%%%%%%%%%%%%%%%%%%%%%%%%%%%%
% You may repeat \author \address as often as necessary      %
%%%%%%%%%%%%%%%%%%%%%%%%%%%%%%%%%%%%%%%%%%%%%%%%%%%%%%%%%%%%%%

\maketitle

\abstracts{
New data on neutrino masses is very exciting.  It is the first evidence for physics beyond the standard model.  In this talk we discuss this data and how it may (or may not) be used to constrain theories of charged fermion masses.  The talk is divided into two parts.  In part I, we briefly discuss the theoretical framework for fermion masses and mixing angles as well as some of the most salient data for neutrino oscillations.  In part II, we define the  framework of a "predictive" theory of fermion masses.  We then consider a particular theory which makes significant testable predictions.}

\section{Introduction to Neutrino Oscillations}
\subsection{Theory}\label{subsec:theory}
Consider the fermion mass sector of the minimal supersymmetric standard model[MSSM].  Fermion masses are defined in the superpotential [$W$]. Three 3 x 3 complex Yukawa matrices  $\lambda_u, \lambda_d, \lambda_e$ are needed to describe up, down and charged lepton masses, while $\lambda_\nu$ is needed for neutrino masses.  In the effective low energy theory, the first three matrices contain 13 observable parameters - 9 masses and 4 quark mixing angles.  The superspace potential is given by
\begin{eqnarray}
W \supset & q_0 \, \lambda_u \, \bar u_0 \, H_u  & + q_0 \, \lambda_d \, \bar d_0 \ H_d \nonumber \\ 
&  + l_0 \, \lambda_\nu \, \bar \nu_0 \, H_u  & + l_0 \, \lambda_e \, \bar e_0  \, H_d  \\
&  +  \frac{1}{2} \; \bar \nu_0^T \, M \, \bar \nu_0 &  \nonumber 
\end{eqnarray}
where all fields are described by two-component left-handed Weyl spinors.  The two Higgs doublets $H_u, \, H_d$ couple to the quark and lepton doublets 
\begin{eqnarray}  q & = \left( \begin{array}{c} u \\ d \end{array} \right) & \\
                  l & = \left( \begin{array}{c} \nu \\ e \end{array} \right) & . \nonumber 
\end{eqnarray}
and the left-handed anti-quark $\bar u, \, \bar d$ and anti-lepton $\bar e, \, \bar \nu$ singlets.
Finally, we have added three sterile neutrinos $\bar \nu$ into the MSSM in order to describe neutrino masses and mixing angles.  They are sterile since they have no electroweak quantum numbers.   As a result, a 3 x 3 Majorana mass term with mass matrix $M$ may be added to the theory without breaking any local gauge symmetry.  When the eigenvalues of $M$ are taken much larger than the weak scale we obtain the well-known see-saw mechanism.\cite{grsy}  We return to neutrino masses shortly, but first consider the charged fermion masses.

In a fundamental theory, presumably defined at some large scale, the quark and lepton Yukawa matrices are complex 3 x 3 matrices which encode fermion masses and mixing angles.  In this basis, labeled by the subscript $0$, the Yukawa matrices are not diagonal; while electroweak interactions are diagonal.  This is the so-called weak eigenstate basis.  At the weak scale the electrically neutral Higgs components $H_u^0, H_d^0$ obtain vacuum expectation values[vev] $\frac{v_u}{\sqrt{2}} = \frac{v \, \sin\beta}{\sqrt{2}} $, $\frac{v_d}{\sqrt{2}} = \frac{v \, \cos\beta}{\sqrt{2}} $, respectively, where  $v = (\sqrt{2} G_F)^{-1/2} = 246$ GeV and $G_F$ is the Fermi constant. We then identify the charged fermion mass matrices  $m_u = \lambda_u \frac{v_u}{\sqrt{2}}$ and 
similarly for down and charged lepton masses.  The $u, d, e$ mass matrices can each be diagonalized by two specific unitary matrices.  For example, for up quarks we have 
\begin{eqnarray}
\left( \begin{array}{ccc} m_u &  &  \\  & m_c &  \\  &  & m_t \end{array} \right) &  = U_{L u}^\dagger  \, m_u \, U_{R u} . &  
\end{eqnarray}
In this mass eigenstate basis, electroweak interactions of quarks are no longer flavor diagonal.   Moreover, the CKM matrix is given by the mismatch between diagonalizing left-handed up and down quarks.  We have
\begin{eqnarray}
V_{CKM} & \equiv  U_{L u}^\dagger \, U_{L d} &
\end{eqnarray}

Now consider leptons.  The charged lepton mass is given by
\begin{eqnarray}
m_e   &  =   \lambda_e  \frac{v_d}{\sqrt{2}}  &
\end{eqnarray}
and is diagonalized by the transformation  $l_0 = l \, U^\dagger_{L e}$  and
similarly for left-handed anti-leptons such that
\begin{eqnarray}
e_0 \, m_e \,\bar e_0   &  = e \, U_{L e}^\dagger \, m_e \, U_{R e} \, \bar e &  \\
& = e  \left( \begin{array}{ccc} m_e &  &  \\  & m_\mu &  \\  &  & m_\tau \end{array} \right) \bar e &  \nonumber
\end{eqnarray}

For neutrinos we now have
\begin{eqnarray}
  &   l \, U_{L e}^\dagger \, \lambda_\nu \, \bar \nu_0 \, H_u  &  +  \frac{1}{2} \, \bar \nu_0^T \, M \, \bar \nu_0 
\end{eqnarray}
The mass scale $M$ may be much larger than the weak scale; thus in the limit
$M >> M_Z$ we can integrate the sterile neutrinos $\bar \nu_0$ out of the superspace potential and obtain the effective dimension five operator, see figure 1 -
\begin{figure}
\vspace{-1cm}
	\centerline{ \psfig{file=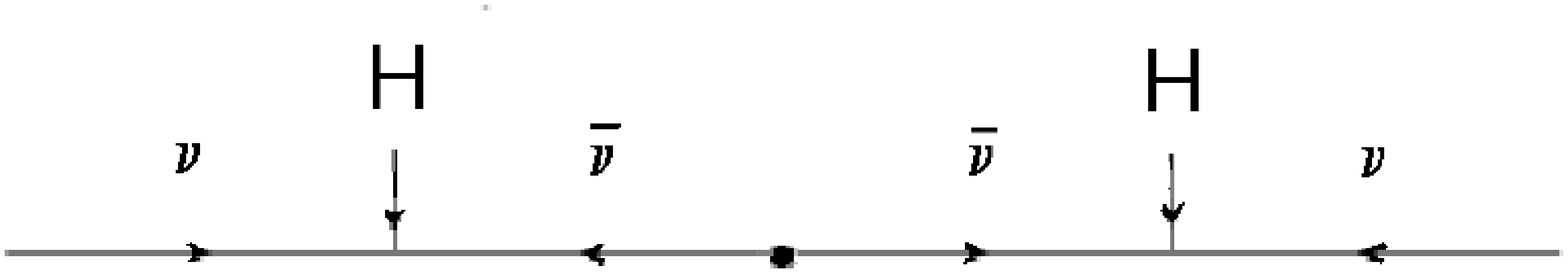,width=10cm,height=3cm}}
\caption{Diagram generating the effective neutrino mass operator}
\end{figure}

\begin{eqnarray}
  &   ( l \, U_{L e}^\dagger \, \lambda_\nu \, H_u ) \, M^{-1} \, ( H_u \, \lambda^T_\nu \, U_{L e}^* \, l ) & 
\end{eqnarray}
This is the Gell-Mann, Ramond, Slansky and Yanagida\cite{grsy} seesaw mechanism for obtaining light Majorana neutrinos.   

The dimension-5 operator is defined at the scale $M$ and must be renormalized down to $M_Z$ where $H_u$ obtains a vev.  The Dirac neutrino mass is defined by  $m_\nu = \lambda_\nu \frac{v_u}{\sqrt{2}}$ and  finally the effective Majorana neutrino mass is given by
\begin{eqnarray}
  m_\nu^{eff} & =  U_{L e}^\dagger \, m_\nu  \, M^{-1} \, m^T_\nu \, U_{L e}^*  &
\end{eqnarray}
It is defined in the basis in which charged lepton masses are diagonal.  This
is the so-called flavor basis for neutrinos.  It is in this basis that a $W$ boson takes an electron into an electron neutrino, etc.     $m_\nu^{eff}$ however is not necessarily diagonal.   It can be diagonalized by a unitary transformation such that
\begin{eqnarray}
\left( \begin{array}{ccc} m_1 &  &  \\  & m_2 &  \\  &  & m_3 \end{array} \right) & =  U^\dagger \, m_\nu^{eff} \, U^*  &
\end{eqnarray}

The unitary matrix  $ U_{\alpha i} $ and mass eigenvalues $m_i$
are the observables in neutrino oscillation experiments.  The indices  $\alpha = \{ e, \, \mu, \, \tau \}$ label the neutrino flavors, while $i = \{ 1, \, 2, \, 3 \}$ label the mass eigenstates.

So consider a process in which lepton $\alpha$ turns (via weak W exchange) into neutrino flavor $\nu_\alpha$, a linear superposition $\sum_i U_{i \alpha} \nu_i(0)$ of neutrino mass eigenstates,  at position $0$.  A distance $L$ away the neutrino   $\nu_i(L) = exp(i m^2_i L/2 E) \nu_i(0)$  is now detected as a linear superposition of neutrinos of flavor $\beta$, see figure 2. 
\begin{figure}
\vspace{-1cm}
	\centerline{ \psfig{file=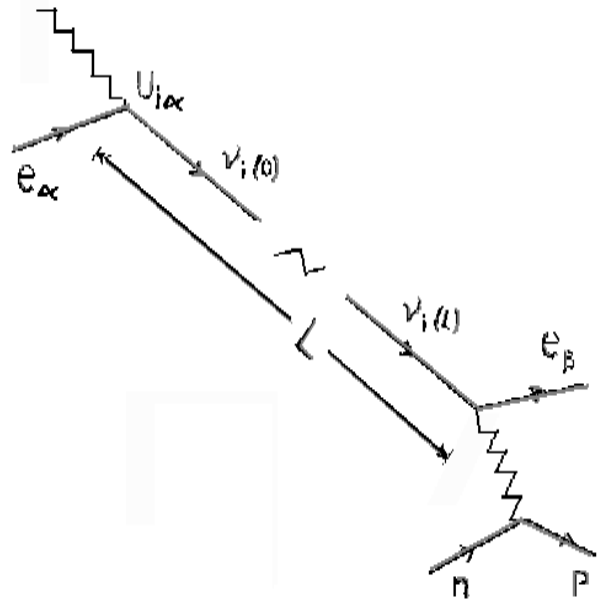,width=10cm,height=10cm}}
\caption{Neutrino of type $\alpha$ produced at point {\bf 0} and detected as neutrino of type $\beta$ a distance {\bf L} away}
\end{figure}
This is the process of vacuum oscillation and the probability for the state $\nu_\alpha$ to remain $\nu_\alpha$ is given by the formula
\begin{eqnarray}
        P(\nu_\alpha \rightarrow \nu_\alpha) & = 1 - 4 \sum_{k < j} P_{\alpha j} P_{\alpha k} \sin^2 \Delta_{j k}   &  \nonumber \\
& &
\end{eqnarray}
where
\begin{eqnarray} 
P_{\alpha j} & \equiv  | U_{\alpha j} |^2  &  \\
\Delta_{j k} & \equiv  \frac{\delta m^2_{j k} L}{4 E} & = 1.27 \left(\frac{\delta m^2_{j k}}{eV^2}\right) \frac{(L/meters)}{(E/MeV)} \nonumber \\   \delta m^2_{j k} & = m_j^2 - m_k^2   &\nonumber 
\end{eqnarray}

Note, when neutrinos travel through matter there are possible matter enhanced neutrino oscillations which were discovered by Mikheyev-Smirnov and Wolfenstein [MSW].\cite{msw}  These effects would take more time to describe and are beyond the scope of this talk.  In what follows, we shall specifically state when MSW oscillations are the dominant effect.

\subsection{Neutrino Oscillation Data}\label{subsec:data}
We briefly discuss the evidence for atmospheric\cite{atm}, solar\cite{solar} and liquid scintillation neutrino detector [LSND]\cite{lsnd} neutrino oscillations.

\vspace{.2in}
\noindent
{\bf Atmospheric neutrino oscillation data}
\vspace{.2in}

Cosmic rays incident on the upper atmosphere scatter on heavy nuclei and produce neutrinos.  The neutrinos are predominantly by-products of pion and kaon decays which result in two muon type neutrinos for each electron neutrino.  These neutrinos are detected at Super-Kamiokande [SuperK]\cite{atm}, a large water cerenkov detector located in a deep undergound laboratory in Japan.  Electron and muon neutrinos scatter in the water producing electrons and muons which are distinguished by their cerenkov rings.    Muons produce a sharp well-defined ring, while electrons multiple scatter in the water producing a ring with fuzzy boundaries.   These atmospheric neutrinos are produced uniformly around the globe and the distance they travel to the detector is a function of the zenith angle $\psi$.   Neutrinos coming from above ($\psi = 0$) typically travel a distance  $L \sim 10$ km, while neutrinos from below the detector ($\psi = \pi$) must travel through the earth corresponding to $L \sim 10,000$ km.   Thus atmospheric neutrinos are a natural source for testing neutrino oscillations.

Experimentalists typically analyze their data in terms of a two neutrino mixing model with mixing matrix  
\begin{eqnarray}
U_{\alpha i}  & =  \left( \begin{array}{cc}  cos\theta & sin\theta \\ -sin\theta & cos\theta \end{array} \right)  & 
\end{eqnarray}
 For muon neutrino oscillations $\alpha = \{\mu, \, x \}$ where $x$ can be either $e, \, \tau$ or $s$ (a sterile neutrino).  The persistence probability for $\nu_\mu$ is given by
\begin{eqnarray}
        P(\nu_\mu \rightarrow \nu_\mu) & = 1 - \sin^2 2\theta \sin^2(\frac{\delta m^2 L}{4 E})  &
\end{eqnarray}
where the muon neutrino energy $E$ is inferred by measuring the energy of the  muon it produces.

The SuperK data~\cite{atm} show (see figure 3)
\begin{itemize}
\item  \# $\nu_\mu$ depends on the zenith angle $\psi$ or equivalently on $L/E$,
\item  \# $\nu_e$ is independent of $L/E$.
\end{itemize}
This simple observation suggests that muon neutrinos oscillate while electron neutrinos do not; hence $\nu_x \neq \nu_e$.
\begin{figure}
\vspace{-1cm}
	\centerline{ \psfig{file=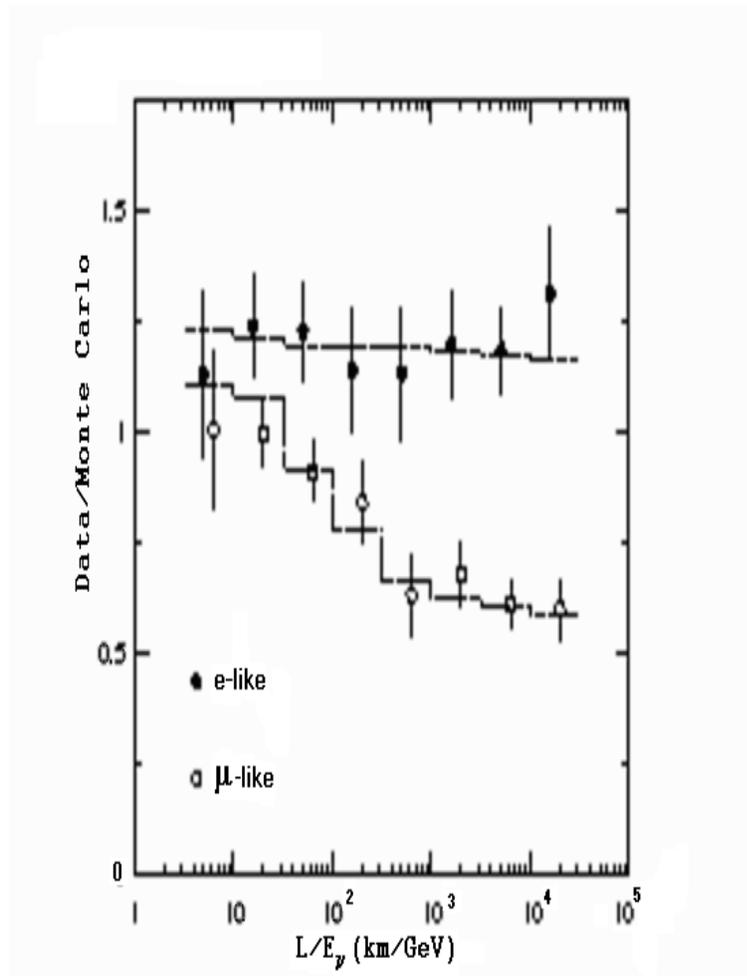,width=10cm,height=13cm}}
\caption{Super-Kamiokande data for electron and muon neutrinos divided by Monte Carlo generated data assuming no oscillations plotted vs. L/E(km/GeV).}
\end{figure}
After 535 days of running the best fit\cite{atm}, see figure 4\cite{atm}, assuming ($\nu_\mu \rightarrow \nu_\tau$)  oscillations (including the fully contained, partially contained and upward going muon data) is given by\footnote{More recent SuperK data\cite{scholberg} after 736 days of fully contained events and 685 days of partially contained events now gives a best fit value of $\delta m^2 = 3.2 \times 10^{-3} \rm eV^2 \rm and \sin^2 2\theta  = 1.05$}
\begin{eqnarray} 
         \sin^2 2\theta & = 1.05 & \\
        \delta m^2 & = 2.2 \times 10^{-3} \; \rm eV^2 &   \nonumber 
\end{eqnarray}
and $\chi^2/d.o.f.  \approx  64.8/67$.  We'll come back to the question of whether or not it makes sense to consider unphysical values of $sin^2 2\theta \ge 1$.
\begin{figure}
\vspace{-1cm}
	\centerline{ \psfig{file=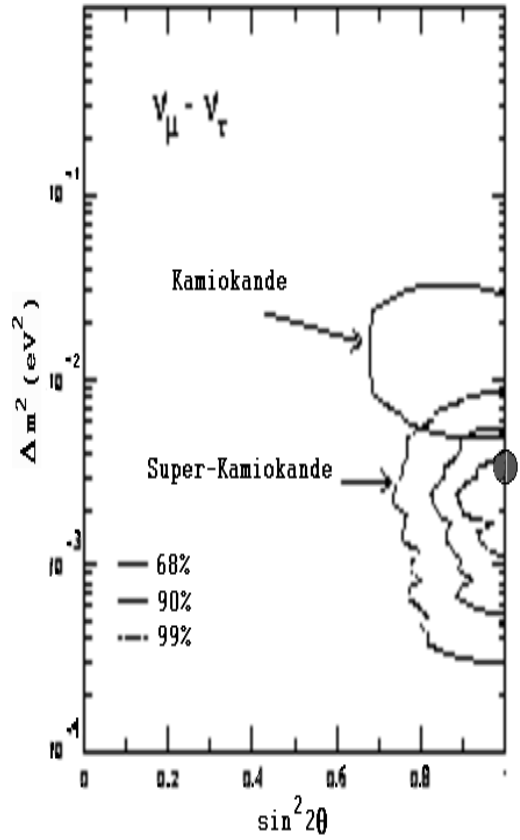,width=10cm,height=13cm}}
\caption{Super-Kamiokande fit to atmospheric neutrino oscillations assuming $\nu_\mu \rightarrow \nu_\tau$.  The figure is taken from SuperK 98.  The large dot corresponds to the result after $\sim$ 700 days of running.}
\end{figure}
If instead ($\nu_\mu \rightarrow \nu_e$) oscillations are assumed, the fit becomes worse with  $\chi^2/d.o.f \approx  87.8/67$.  

In fact, there is additional experimental data which also suggests that $\nu_x \neq \nu_e$.  This comes from the Chooz experiment\cite{chooz}, a neutrino detector which sits about 1 km from a nuclear reactor in France.   The reactor produces electron neutrinos with a known flux.  The experiment looks for the disappearance of these neutrinos.  It is sensitive to the region of parameter space with $\delta m^2 \ge 10^{-3} \rm eV^2$ and $sin^2 2\theta \sim 1$.  Their null result suggests that muon neutrinos cannot oscillate into electron neutrinos with these parameters, since otherwise they would have observed a disappearance of electron neutrinos due to their oscillation into muon neutrinos.    

Thus it is safe to conclude that  $\nu_x \neq \nu_e$ for atmospheric neutrino oscillations.  Hence we have two alternatives,  $\nu_x = \nu_\tau$ or $\nu_s$.  How can this be reconciled?

 We consider a recent analysis by K. Scholberg at SuperK.\cite{scholberg}    At the moment both possibilities fit the zenith angle dependence well.  It should however be possible with better statistics to distinguish $\nu_\tau$ vs. $\nu_s$ by the zenith angle dependence.  This is because there is an MSW effect in the earth for  $\nu_s$ but not for $\nu_\tau$.

Also by considering the ratio of neutral current [NC] to charged current [CC] processes one can distinguish between the two.   In this case there is preliminary data which favors $\nu_x = \nu_\tau$.   
This ratio satisfies
 \begin{eqnarray}
R_{(NC/CC)}  & < 1 & \;\; \rm for \; \nu_x = \nu_s  \label{eq:R1} \\ 
             & = 1  & \;\; \rm for \; \nu_x = \nu_\tau  . \nonumber 
\end{eqnarray} 
Using SuperK data for $\pi^0$ events produced by neutral current neutrino scattering in the detector one measures (see figure 5)
\begin{figure}
\vspace{-1cm}
	\centerline{ \psfig{file=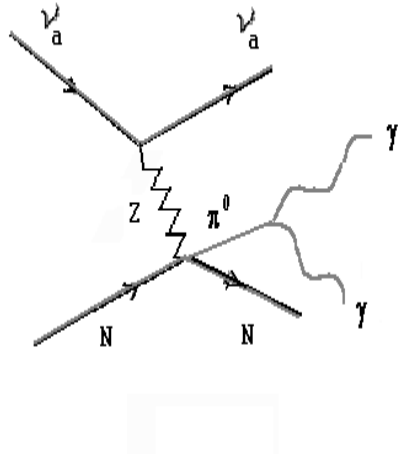,width=10cm,height=7cm}}
\caption{Neutral current process resulting in a $\pi^0 \rightarrow 2 \gamma$ where the subscript a refers to any active neutrino.}
\end{figure}

\begin{eqnarray}
R_{(NC/CC)}  & \equiv  \frac{(\pi^0/e)_{Data}}{(\pi^0/e)_{MonteCarlo}} & \label{eq:R2} \\
&  =  1.11  \pm  0.06 ({\rm data \; stat.}) \pm 0.02 ({\rm MC \; stat.}) \pm 0.26 ({\rm sys.}) & \nonumber 
\end{eqnarray}

\vspace{.2in}
\noindent
{\bf Solar neutrino oscillation data}
\vspace{.2in}

The flux of solar electron neutrinos from the sun can be calculated, see for example.\cite{bp98}
The result depends on the details of the solar model, but independent of which solar model is employed one finds disagreement with the data (see table \ref{tab:sndata}).\footnote{Note the most uncertain component of the neutrino flux is the highest energy so-called hep neutrinos.  This affects the prediction for the high energy tail of solar neutrinos.}

\begin{table}[t]
\caption{Solar neutrino data\label{tab:sndata}}
\begin{center}
\footnotesize
\begin{tabular}{|l|c|}
\hline
  &  Result\cite{solar}/Theory\cite{bp98}  \\
\hline
Homestake&  $.33 \pm  .029$\\
Kamiokande&  $.54 \pm .07$ \\
GALLEX &  $.60 \pm  .06$ \\
SAGE &  $.52  \pm  .06 $ \\
SuperK &  $.474 \pm  .020 $ \\
\hline
\end{tabular}
\end{center}
\end{table}

Solar neutrino data is typically analyzed in terms of two neutrino oscillations where $\nu_e \rightarrow \nu_x$ with $x = a$, an active neutrino ($\nu_\mu, \, \nu_\tau$) or $x = s$, a sterile neutrino.    The results of many fits to all the data give three possible solutions with $x = a$ and one solution with $x = s$ (see table \ref{tab:solarfit}).\cite{bks}   These solutions fall into three categories --- (1) small mixing angle [SMA]  MSW,  (2) large mixing angle [LMA] MSW and (3) the so-called "Just so" vacuum oscillation solution.

\begin{table}[t]
\caption{Solar neutrino fits\label{tab:solarfit}}
\begin{center}
\footnotesize
\begin{tabular}{|l|lc|}
\hline
 MSW  & SMA  &  $sin^2 2\theta \sim 10^{-3} - 10^{-2} $ \\
    & $\nu_x = \nu_s, \, \nu_a$  &  $\delta m^2 \sim 10^{-6} - 10^{-5}$ eV$^2$\\
\hline
      &  LMA   & $sin^2 2\theta \sim .6 - .9$ \\
      & $\nu_x = \nu_a$  &  $\delta m^2 \sim 10^{-5} - 10^{-4}$ eV$^2$ \\
\hline
Vacuum  &   "Just so"  &  $sin^2 2\theta \sim 1 $ \\
    & $\nu_x = \nu_a$  &  $\delta m^2 \sim 10^{-11} - 10^{-10}$ eV$^2$\\
\hline
\end{tabular}
\end{center}
\end{table}

 \vspace{.2in}
\noindent
{\bf LSND oscillation data}
\vspace{.2in}

The LSND experiment\cite{lsnd} uses a proton beam at Los Alamos National Lab to produce pions in a beam dump.   Both $\pi^+$s and $\pi^-$s are produced.   In their stopped pion data,  $\pi^+$s decay at rest into $\mu^+  \nu_\mu$ and $\mu^+ \rightarrow e^+ \;  \nu_e  \; \bar \nu_\mu$.   Note, there are no $\bar \nu_e$s in this decay chain.   $\pi^-$s, on the otherhand will typically be absorbed into a nucleus before they have a chance to decay.

They then look for the signature of $\bar \nu_e$ in the detector, see figure 6.
  The $\bar \nu_e$ scatters on a proton and produces $e^+ \, n$ in the detector.  The positron is observed via its cerenkov radiation.  A short time later the neutron finds a nucleus and is captured, producing a gamma ray.  The coincidence of these two signals is required for an event.  This is the only appearance experiment and they find a signal.    Once again the signal is parametrized in terms of two neutrino mixing.   They find 
\begin{eqnarray} 
         \sin^2 2\theta & \sim 10^{-3} - 3\times 10^{-2} & \\
        \delta m^2 & \sim  .2 - 2 \; \rm eV^2 &   \nonumber 
\end{eqnarray}

\begin{figure}
\vspace{-1cm}
	\centerline{ \psfig{file=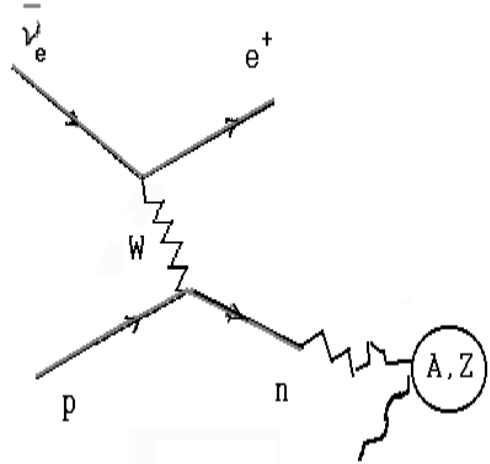,width=10cm,height=7cm}}
\caption{LSND detects an anti-electron neutrino via the correlated detection of positron cerenkov radiation and neutron capture.}
\end{figure}

\vspace{.2in}
\noindent
{\bf Summarizing neutrino oscillation data}
\vspace{.2in}

There are some general conclusions we may draw from the above data.    
\begin{figure}
\vspace{-1cm}
	\centerline{ \psfig{file=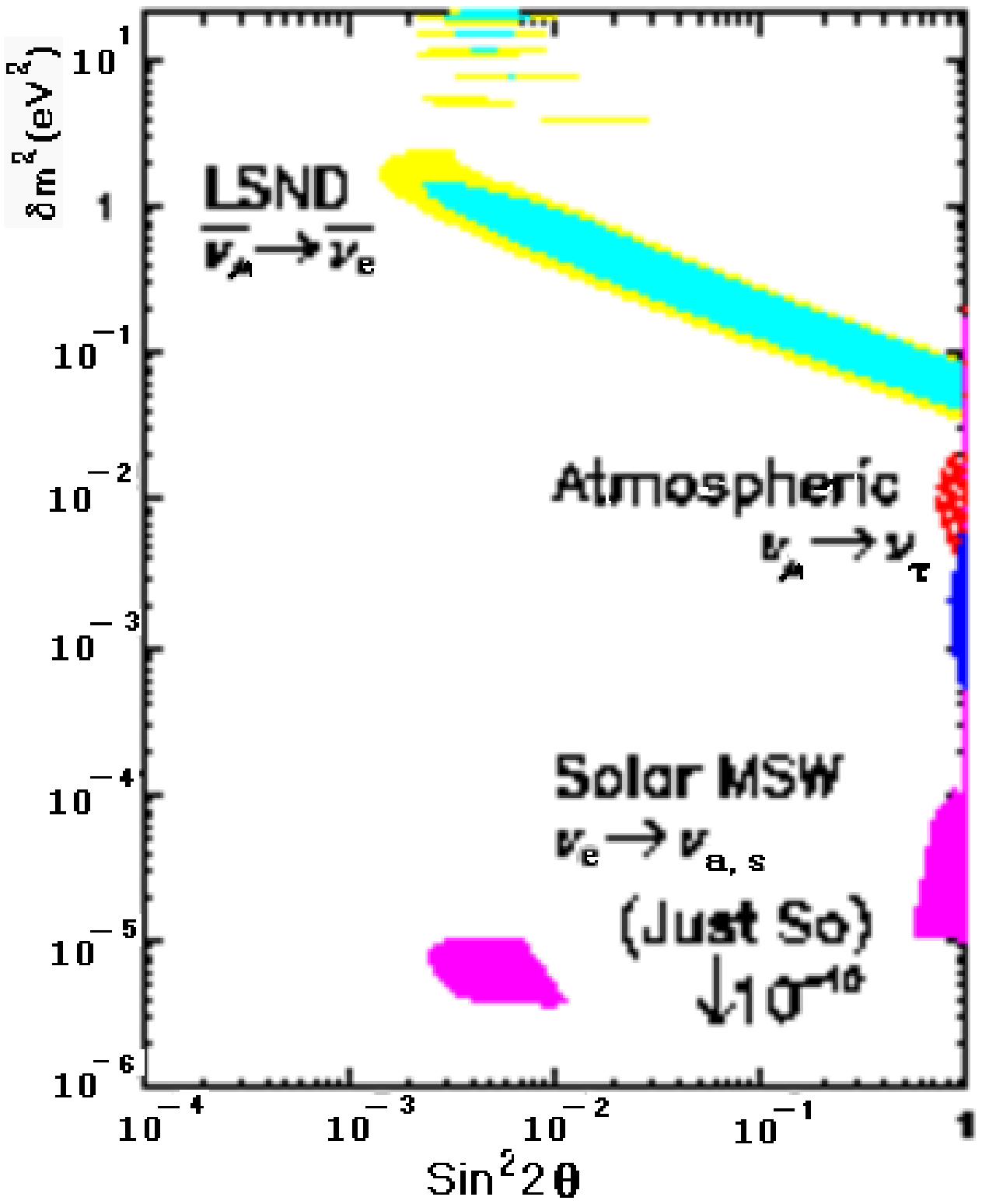,width=10cm,height=14cm}}
\caption{Summary of neutrino oscillation data}
\end{figure}

The data give three distinct scales for neutrino oscillations, see figure 7.\cite{conrad}   

We see
 \begin{eqnarray} 
        \delta m^2_{LSND} & >>  \delta m^2_{Atm.}  &   >>  \delta m^2_{Solar} .
\end{eqnarray}

$\bullet$ {\em Thus IF all three experiments are correct, we necessarily need four light neutrinos.}  This of course requires {\em sterile neutrinos}.
 
Atmospheric neutrino oscillations, and perhaps even solar neutrinos, require maximal mixing ($sin^2 2\theta \sim 1$), while quark mixing angles are all small.       We thus have
\begin{eqnarray} 
        U_{\alpha i} & >>  V_{CKM}  &   .
\end{eqnarray}
$\bullet$  {\em Hence any theory of fermion masses must give large mixing for leptons and small mixing for quarks.  This is especially difficult in grand unified theories which relate quarks and leptons.}

\section{"Predictive" SUSY GUT}\label{sec:predictive}
At this point let me define the framework for theories of fermion masses we call a "predictive" theory.  
\begin{enumerate}
\item  It is a "natural" theory, i.e. if one gives you the symmetries of the theory and also the states and their transformations under these symmetries, then you can write down the Lagrangian.    It is the most general Lagrangian consistent with the symmetries.   However, this is not sufficient.  
\item  We also require that the number of arbitrary parameters in the theory is less than the number of observables.
\end{enumerate}
A predictive theory is testable.

Clearly, to have a predictive theory one must necessarily have lots of symmetry in order to reduce the number of arbitrary parameters.   The appropriate symmetry is determined by the data.

\vspace{.2in}
\noindent
{\bf SUSY GUTs}
\vspace{.2in}

Consider first supersymmetric grand unification or {\bf SUSY GUTs} .\cite{susyguts}  SUSY dramatically restricts the form of the Lagrangian, while GUTs unify quarks and leptons;  thus quark and lepton Yukawa matrices may be related by Clebsch-Gordon coefficients.  The group may be $SU_5,\, SO_{10}$ or any other simple group which breaks to the standard model gauge group at a high energy scale  $M_G$.   Such a theory is known to be consistent with precision electroweak data.   For example, given the observed values for $\alpha_1, \, \alpha_2$ at $M_Z$ one finds that the running couplings unify at a scale $M_G \sim 2\times 10^{16}$ GeV.  Running the value of $\alpha_3$ down from $M_G$ to $M_Z$ one predicts the value of the strong coupling at the Z scale which agrees quite well with the data.

\vspace{.2in}
\noindent
{$\bf SO_{10}$}
\vspace{.2in}

We use the GUT symmetry $SO_{10}$.\cite{SO(10)}  This is the minimal symmetry in which all fermions in one family are contained in a single irreducible representation.  We have\footnote{ Note, $\bar \nu$ denotes a sterile neutrino.}
\bea
16 & \supset \{ q, \, \bar u, \, \bar d; \, l, \, \bar e, \, \bar \nu \} &
\eea

In the simplest version of $SO_{10}$ one pair of Higgs doublets are contained in a $10$.  The minimal Yukawa coupling of one family of fermions to the Higgs fields is given by  $ \lambda \; 16 \, 10 \, 16 $ where $\lambda$ is the single Yukawa coupling.   If we use this minimal coupling for the heaviest generation we obtain the symmetry relation $\lambda_t = \lambda_b = \lambda_\tau = \lambda_{\nu_{\tau}} = \lambda$ at the GUT scale.  This is Yukawa unification and experimentally it is known to work quite well for the third generation\cite{yukawaunification}, but fails miserably for the other two.  Using the values of $m_b(m_b), \, m_\tau(m_\tau)$, one determines both  $\lambda(M_G)$ and $\tan\beta$.  One then predicts the value of the top quark mass\cite{yukawaunification,halletal}
\bea
m_t(m_t) & = \lambda_t(m_t) \frac{v}{\sqrt{2}} \sin\beta & \sim 170 \pm 20 \;\; \rm GeV .
\eea

A brief digression: We note that the success of Yukawa unification has important ramifications for string theory model building.   String theories, with SUSY GUTs realized as level one Kac-Moody algebras, necessarily need Wilson line breaking of the SUSY GUT down to the standard model gauge group.  Unfortunately, Wilson line breaking preserves the nice feature of gauge coupling unification but typically destroys the equally nice feature of Yukawa coupling unification.   This is a problem with this realization of GUTs in string theory.

Finally,  whereas $SO_{10}$ unifies all quarks and leptons in a single family into one irreducible representation of the group, relating up to down to charged lepton to Dirac neutrino masses,  it is not sufficient to obtain a "predictive" theory.   We need more symmetry; a family symmetry to relate states of different families and to reduce the number of arbitrary parameters.  Moreover family symmetry breaking can provide an understanding of the hierarchy of fermion masses and also ameliorate problems with flavor violating processes in SUSY theories.\cite{flavorviolation}  To summarize, family symmetry plus GUTs are necessary for a "predictive" theory, but they are not necessarily sufficient.  It is still necessary to check in each case whether or not they sufficiently constrain the theory.

\subsection{$U_2 \times U_1$ family symmetry}
In this talk, we consider a particular $SO_{10}$ SUSY GUT with family symmetry {$U_2 \times U_1\times \cdots$.\cite{brt}   This model is a slight variation of the one considered previously in a paper by Barbieri et al. [BHRR].\cite{bhrr}  The changes affect the treatment of neutrinos without affecting the predictions for charged fermion masses.   

The three families transform as a doublet  $16_a,  \; a = 1,2$ and a singlet $16_3$ under $U_2$ with $U_1$ charge (-1), (0), respectively.  Note, the $U_1$ contained in $U_2$ counts +1 (-1) for every upper (lower) $SU_2$ index.  The hierarchy of fermion masses is explained by the breaking of the family symmetry.   At tree level only the third family, $16_3$, can couple to the $10$ of Higgs bosons which also carries no family symmetry charge.  There are three "flavon" fields in the theory $(\phi^a ,\; S^{a b} = S^{b a}, \; A^{a b} = - A^{b a})$ transforming (under $SU_2 \times U_1 \times U_1$) as (2,1,0), (3,2,1), (1,2,2) responsible for spontaneously breaking the family symmetry.   The vacuum expectation values [vevs] $<\phi^2> \approx <S^{2 2}>$ spontaneously break $U_2 \times U_1$ to $\tilde U_1$ and give the second family mass.  Then $<A^{1 2}>$ breaks the remaining symmetry allowing the first family to obtain mass.\footnote{It is important to have only these flavon vevs in order to retain a predictive theory.}

It is important to recall that the non-abelian $SU_2$ family symmetry can also protect against large flavor violations such as $\mu \rightarrow e \gamma$.\cite{u2symmetry}

\subsection{The "predictive" theory}
This theory is completely defined by the symmetries, in this case
\bea  SO_{10}  \times & [\; U_2 \times U_1 \times U_1(R) \times U_1(PQ) \times \cdots \;] &   \\
   &  {\rm family \;\;\; symmetry} & \nonumber \eea
where $U_1(R), \, U_1(PQ)$ denote an R symmetry for which all fields, including $M, \, M', \, M$'' have charge +1 and a Peccei-Quinn symmetry for which all $16$s have charge +1.   The ellipsis denotes additional symmetries which exist in the fermion sector of the theory which may or may not remain unbroken in the complete theory.\footnote{One of these symmetries is needed to distinguish the fields $M$ and $M^\prime, \, M^{\prime \prime}$.}

Given the states in the theory and their charges we obtain the superspace potential $W$ given by (the $U_1$ charge is denoted by a superscript)
\bea
W & =  W_{charged \, fermion} + W_{neutrino} +  \; \cdots & 
\eea
with
\bea
W_{charged \, fermion} =&  16_3^{\bf 0} \, 10^{\bf 0} \, 16_3^{\bf 0} & + 16_a^{\bf -1} \, 10^{\bf 0} \, (\chi^a)^{\bf 1} \\
& +  \bar \chi_a^{\bf -1} \, [\, M^{\bf 0} \, (\chi^a) ^{\bf 1} + (\phi^a) ^{\bf 0} \, \chi^{\bf 1} &  +  (S^{a b})^{\bf 1} \, \chi_b^{\bf 0}  +  (A^{a b})^{\bf 2} \, 16_b^{\bf -1} \,] \nonumber \\
& +  (\bar \chi^a) ^{\bf 0} \, [\, M'^{\bf 0}  \, \chi_a^{\bf 0} +  45^{\bf 1} \, 16_a^{\bf -1} ]  & + \bar \chi^{\bf -1} \,[\, M"^{\bf 0}  \, \chi^{\bf 1} +  45^{\bf 1}  \, 16_3^{\bf 0} ] \nonumber
\eea
\bea
W_{neutrino} = &  (\overline{16})^{\bf -1/2} \,[\, N_a^{\bf -1/2} \, (\chi^a) ^{\bf 1} + N_3^{\bf 1/2} \, 16_3^{\bf 0} ] & \\
&  + (S^{a b})^{\bf 1} \, N_a^{\bf -1/2} \, N_b^{\bf -1/2} +  (\phi^a) ^{\bf 0} \, N_a^{\bf -1/2} \, N_3^{\bf 1/2} & \nonumber \label{eq:Wneutrino}
\eea

The vevs of the fields  $M,\, M$', $M$'' give mass to the Froggatt-Nielsen\cite{fn} states labelled by $\chi,\, \bar \chi$.  The fields $M$', $M$'' are $SO_{10}$ singlets and we use the same notation for their vevs.  $M$, on the otherhand, is a linear superposition of an $SO_{10}$ singlet and adjoint with vev $M_0 ( 1 + \alpha_0 \, X + \beta_0 \, Y)$  where $ \alpha_0, \,\beta_0$ are arbitrary constants which are fit to the data and $X,\,Y$ are elements of the Lie algebra of $SO_{10}$ with $X$ in the direction of the $U_1$ which commutes with $SU_5$ and $Y$ the standard weak hypercharge.  Each term in $W$ has an arbitrary Yukawa coupling which is implicit.   Finally, the fields $N_a, \, N_3$ are $SO_{10}$ singlets; introduced to generate neutrino masses.  These will be discussed in more detail shortly.

The largest scale in the theory is assumed to be the Froggatt-Nielsen masses.  Below this mass scale, the $\chi,\, \bar \chi$ states are integrated out of the theory giving the effective mass operators, see figure 8.   
\begin{figure}
\vspace{-1cm}
	\centerline{ \psfig{file=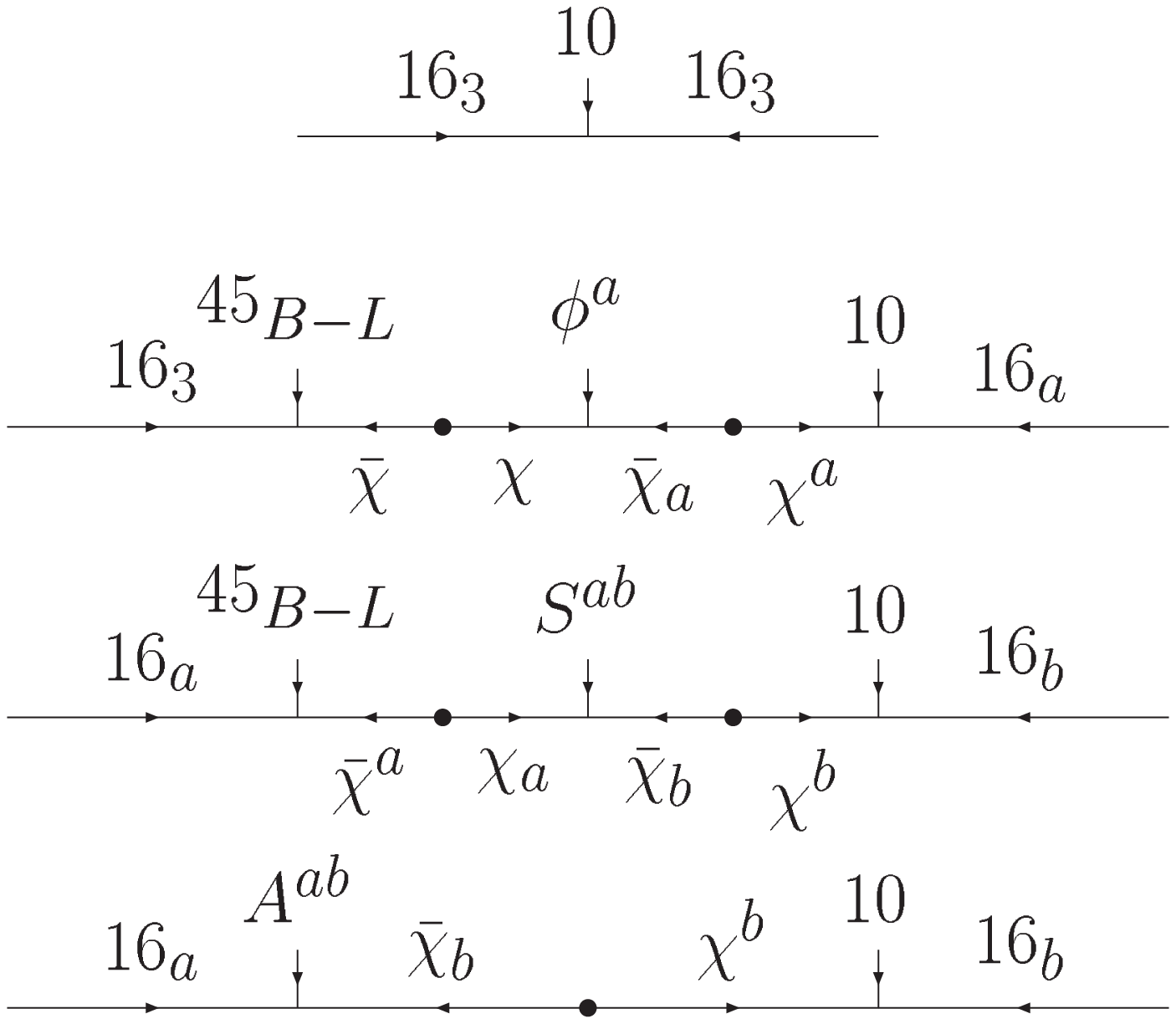,width=10cm,height=21cm}}
\caption{Diagrams generating the Yukawa matrices}
\end{figure}
Finally, when the "flavon" fields obtain vevs 
\bea
<\phi^a>  & = \phi \, \delta^{a 2} &  \\
<S^{a b}>  &  =  S \,  \delta^{a 2} \delta^{b 2}  &  \nonumber \\
<A^{1 2}>  &  \neq 0   & \nonumber 
\eea
we generate the Yukawa couplings at a scale of order the GUT scale given by\footnote{Note the parameters $ \lambda, \, \epsilon, \, \epsilon', \, \rho, \, \sigma, \, r$ are implicit functions of ratios of vevs, Yukawa couplings and $\alpha_0,\, \beta_0$.}
\begin{eqnarray}
Y_u =&  \left(\begin{array}{ccc}  0 & \epsilon' \rho & 0 \\
                          - \epsilon' \rho &  \epsilon \rho & r \epsilon
T_{\bar
u}     \\
                      0  & r \epsilon T_Q& 1 \end{array} \right) \; \lambda &
\nonumber \\
Y_d =&  \left(\begin{array}{ccc}  0 & \epsilon'  & 0 \\
                          - \epsilon'  &  \epsilon  &  r \sigma \epsilon
T_{\bar
d}\\
                      0  &  r \epsilon T_Q & 1 \end{array} \right) \; \lambda & \label{eq:yukawa}
  \\
Y_e =&  \left(\begin{array}{ccc}  0 & - \epsilon'  & 0 \\
                           \epsilon'  &  3 \epsilon  &  r \epsilon T_{\bar
e} \\
                      0  &  r \sigma \epsilon T_L & 1 \end{array} \right)
\; \lambda &
 \nonumber \\
Y_{\nu} =&  \left(\begin{array}{ccc}  0 & - \omega \epsilon'  & 0 \\
                  \omega \epsilon'  &  3 \omega \epsilon  & {1 \over 2}
\omega r
\epsilon T_{\bar \nu}
\\
                      0  &  r \sigma \epsilon T_L& 1 \end{array} \right) \;
\lambda &
 \nonumber
\end{eqnarray}
with  \begin{eqnarray} \omega = {2 \, \sigma \over 2 \, \sigma - 1}
\label{eq:omega} \end{eqnarray} and
\begin{eqnarray} T_f  = & (\rm Baryon\# - Lepton \#) &
\label{eq:Tf} \\
\rm for & f = \{Q,\bar u,\bar d, L,\bar e, \bar \nu\}.& \nonumber
\end{eqnarray}

There are six real parameters in these matrices $ \lambda, \, \epsilon, \, \epsilon', \, \rho, \, \sigma, \, r$ and three phases which cannot be rotated away, which we take to be $\Phi_\epsilon, \, \Phi_\sigma, \, \Phi_\rho$.   These nine parameters are then fit to the thirteen observable charged fermion masses and quark mixing angles.   Note, once these parameters are determined the Dirac neutrino mass matrix $m_\nu$ is fixed.  Before we consider neutrino masses and mixing, we should first test our theory against precision electroweak data which, in our case, includes the data on charged fermion masses and mixing angles.  We shall see that the fits are quite good.

\subsection{Results for Charged Fermion Masses and Mixing Angles}

\protect
\begin{table}
\caption[8]{
{\bf Charged fermion masses and mixing angles} \\
   \mbox{Initial parameters:}\ \ \
\ \

 (1/$\alpha_G, \, M_G, \, \epsilon_3$) = ($24.52, \, 3.05 \cdot 10^{16}$
GeV,$\,
-4.08$\%), \makebox[1.8em]{ }\\
 ($\lambda, \,$r$, \, \sigma, \, \epsilon, \, \rho, \, \epsilon^\prime$) =
($ 0.79, \,
12.4, \, 0.84, \, 0.011, \,  0.043,\,  0.0031$),\\
($\Phi_\sigma, \, \Phi_\epsilon, \, \Phi_\rho$) =  ($0.73, \, -1.21, \,
3.72$)rad,
\makebox[6.6em]{ }\\
($m_0, \, M_{1/2}, \, A_0, \, \mu(M_Z)$) = ($1000,\, 300, \, -1437, \,
110$) GeV,\\
($(m_{H_d}/m_0)^2, \, (m_{H_u}/m_0)^2, \, $tan$\beta$) = ($2.22,\, 1.65, \,
53.7$)
% end of caption
}
\label{t:fit4nu}
$$
\begin{array}{|l|c|l|}
\hline
{\rm Observable}  &{\rm Data}(\sigma) & Theory  \\
\mbox{ }   & {\rm (masses} & {\rm in\  \ GeV) }  \\
\hline
\;\;\;M_Z            &  91.187 \ (0.091)  &  91.17          \\
\;\;\;M_W             &  80.388 \ (0.080)    &  80.40       \\
\;\;\;G_{\mu}\cdot 10^5   &  1.1664 \ (0.0012) &  1.166     \\
\;\;\;\alpha_{EM}^{-1} &  137.04 \ (0.14)  &  137.0         \\
\;\;\;\alpha_s(M_Z)    &  0.1190 \ (0.003)   &  0.1174       \\
\;\;\;\rho_{new}\cdot 10^3  & -1.20 \ (1.3) & +0.320   \\
\hline
\;\;\;M_t              &  173.8 \ (5.0)   &  175.0       \\
\;\;\;m_b(M_b)          &    4.260 \ (0.11)  &    4.328                  \\
\;\;\;M_b - M_c        &    3.400 \ (0.2)   &    3.421                 \\
\;\;\;m_s              &  0.180 \ (0.050)   &  0.148          \\
\;\;\;m_d/m_s          &  0.050 \ (0.015)   &  0.0589        \\
\;\;\;Q^{-2}           &  0.00203 \ (0.00020)  &  0.00201                \\
\;\;\;M_{\tau}         &  1.777 \ (0.0018)   &  1.776         \\
\;\;\;M_{\mu}          & 0.10566 \ (0.00011)   & .1057           \\
\;\;\;M_e \cdot 10^3      &  0.5110 \ (0.00051) &  0.5110  \\
 \;\;\;V_{us}         &  0.2205 \ (0.0026)      &  0.2205        \\
\;\;\;V_{cb}         & 0.03920 \ (0.0030)      &  0.0403           \\
\;\;\;V_{ub}/V_{cb}    &  0.0800 \ (0.02)    &  0.0691                 \\
\;\;\;\hat B_K          &  0.860 \ (0.08)    &  0.8703           \\
\hline
{B(b\!\rightarrow\! s \gamma)\!\cdot\!10^{4}}  &  3.000 \ (0.47) &  2.995  \\
\hline
  \multicolumn{2}{|l}{{\rm TOTAL}\;\;\;\; \chi^2}  3.39
            & \\
\hline
\end{array}
$$
\end{table}

We have performed a global $\chi^2$ analysis, incorporating two (one) loop
renormalization group[RG] running of dimensionless (dimensionful)
parameters from $M_G$ to $M_Z$ in the MSSM,  one loop radiative threshold 
corrections at $M_Z$, and 3 loop QCD (1 loop QED) RG running below
$M_Z$.~\footnote{The predicted values of the low energy observables are highly 
correlated.  Thus a global $\chi^2$ analysis is necessary in order to test the
accuracy of the fit. We note that fermion masses and mixing angles 
are the precision electroweak data which constrain any theory beyond the 
Standard Model.  It is important to know how well a theory beyond the Standard
Model fits this data, even though in some cases this data still has large 
theoretical uncertainties.} 
Electroweak symmetry breaking is obtained self-consistently from the 
effective potential at one loop, with all one loop threshold corrections 
included. This analysis is performed using the code of Blazek
et.al..~\cite{blazek}~\footnote{We assume universal scalar mass $m_0$ for
squarks and sleptons at $M_G$.  We have not considered the flavor violating 
effects of U(2) breaking scalar masses in this work.}  In this work\cite{brt}, 
we present the results for one set of soft SUSY breaking parameters 
$m_0, \; M_{1/2}$ with all other parameters varied to obtain the best 
fit solution.  In table \ref{t:fit4nu} we give the 20 observables which 
enter the $\chi^2$ function, their experimental values and the uncertainty 
$\sigma$ (in parentheses).\footnote{ The Jarlskog parameter $J = 
Im(V_{ud}V_{ub}^*V_{cb}V_{cd}^*)$ is a measure of CP violation.
We test $J$ by a comparison to the experimental value extracted from 
the well-known $K^0-\overline{K^0}$ mixing observable 
$\epsilon_K =  (2.26 \pm 0.02)\times10^{-3}$.  The largest uncertainty in such
a comparison, however, comes in the value of the QCD bag constant $\hat B_K$. 
 We thus exchange the Jarlskog parameter $J$ for $\hat B_K$ in the list 
of low-energy data we are fitting. Our theoretical value of $\hat B_K$ is 
defined as that value needed to agree with $\epsilon_K$
for a set of fermion masses and mixing angles derived from the
GUT-scale. We test this theoretical value against the ``experimental'' 
value of $\hat B_K$. This value, together with its error estimate, is 
obtained from recent lattice calculations.\cite{kilcup}}   In most
cases $\sigma$ is determined by the 1 standard deviation experimental 
uncertainty, however in some cases the theoretical  uncertainty 
($\sim$ 0.1\%) inherent in our renormalization group running and one 
loop threshold corrections dominates.

  Given the 6 real Yukawa parameters and 3 complex phases we fit the 13 fermion mass observables (charged fermion masses and mixing angles [$\hat{B}_K$ replacing $\epsilon_K$ as a ``measure of CP violation"]); we thus
have 4 predictions.  From table \ref{t:fit4nu}  it is clear that this theory fits the low energy data quite well.~\footnote{In a future paper we intend to  explore the dependence of the fits on the SUSY breaking parameters and also U(2) flavor violating effects. Note also the strange quark mass $m_s(1 \rm GeV) \sim 150 \ \rm MeV$ is small, consistent with recent lattice results.  Finally, by adding one new parameter (a small ratio for the $H_d$ to $H_u$ Yukawa couplings) it is possible to obtain a small $\tan\beta$ solution that fits charged fermion masses just as well.}

Finally, the squark, slepton, Higgs and gaugino spectrum of our theory is
consistent with all available data.  The lightest chargino and neutralino are
higgsino-like with the masses close to their respective experimental limits. As an example of the additional predictions of this theory consider the CP
violating mixing angles which may soon be observed at B factories.   For the selected fit we find~\footnote{We warn the reader that according to quite
standard conventions the angle $\beta$ is used in two inequivalent ways.  tan$\beta$ is the ratio of Higgs vevs in the minimal supersymmetric standard model; while $\sin 2\beta$ refers to the CP violating angle in the unitarity triangle, measured in B decays.  We hope that the reader can easily distinguish the two from the context.}
\begin{eqnarray}
(\sin 2\alpha, \, \sin 2\beta, \, \sin \gamma) = & (0.74, \, 0.54, \,
0.99)&
\end{eqnarray}  or equivalently the Wolfenstein parameters
\begin{eqnarray}
(\rho, \, \eta )    =      &( -0.04, \,      0.31)    &.
\end{eqnarray}

\section{Neutrino Masses and Mixing Angles}

The parameters in the  Dirac Yukawa matrix for neutrinos (eqn.
\ref{eq:yukawa}) mixing
$\nu - \bar \nu$ are now fixed.  Of course, neutrino masses are much too
large and we need to invoke the GRSY~\cite{grsy} see-saw mechanism.

Since the {\bf 16} of SO(10) contains the  ``right-handed" neutrinos
$\bar \nu$, one possibility is to obtain  $\bar \nu - \bar \nu$ Majorana masses
via higher dimension operators of the form 
\begin{eqnarray}
{1 \over M} \ \overline{16} \ 16_3 \ \overline{16} \ 16_3   , \\
{1 \over M^2} \ \overline{16} \ 16_3 \ \overline{16} \ 16_a \ \phi^{a}  ,
\nonumber\\ {1 \over M^2} \ \overline{16} \ 16_a \ \overline{16} \ 16_b \
S^{a \, b}
. \nonumber
 \end{eqnarray}
If $\overline{16}$ field gets a vev in the "right-handed" neutrino direction, then we obtain a $\bar \nu - \bar \nu$ mass of order  $<\overline{16}>^2/M$. This possibility
has been considered in the paper by Carone and Hall.\cite{u2neutrino}

The second possibility, which we follow, is to introduce SO(10) singlet
fields $N$ and obtain effective mass terms $\bar \nu - N$ and $N - N$
using only dimension four operators in the superspace potential.  To do this,
we add three new SO(10) singlets
\{$N_a,\; a = 1,2;\;\; N_3$\}  with U(1) charges \{  $-1/2$,\  +1/2 \}.
These then contribute to the superspace potential (see equation \ref{eq:Wneutrino}).
Note the field $\overline{16}$ with U(1) charge  $-1/2$ is assumed to get a
vev in the ``right-handed" neutrino direction; this vev is
also needed to break the rank of SO(10).

Finally in order to allow for the possibility of light sterile neutrinos we introduce a U(2) doublet of SO(10) singlets $\bar N^a$ or a U(2) singlet $\bar N^3$.  They enter the superspace potential as follows --
\begin{eqnarray}
W \supset &  \mu' \; N_a \; \bar N^a \;\; + \;\;\mu_3 \;  N_3 \bar N^3 &
\label{eq:mu'}
\end{eqnarray}
We show that if the dimensionful parameters $\mu', \; \mu_3$ are of order the
weak scale, then the sterile neutrinos (predominantly $\bar N$s) are light.  The notation is suggestive of the similarity between these terms and the $\mu$ term in the Higgs sector. In both cases, we add supersymmetric mass terms and in both cases we need some mechanism to keep these dimensionful parameters small compared to the Planck scale.  This may be accomplished by symmetries, see for example ref. ~\cite{gm}.

We define the 3 $\times $ 3 matrix
\begin{eqnarray}
\tilde \mu = & \left( \begin{array}{ccc}  \mu' & 0 & 0 \\
                               0 & \mu' & 0 \\
                                 0  &  0  & \mu_3 \end{array}\right) &
\end{eqnarray}
The matrix $\tilde \mu$ determines the number of {\em coupled} sterile
neutrinos, i.e.
there are 4 cases labeled by the number of neutrinos ($N_\nu = 3, 4, 5, 6$):
\begin{itemize}
\item ($N_\nu = 3$) \hspace{.1cm} 3 active \hspace{.3cm} ($\mu' = \mu_3 =
0$);
\item ($N_\nu = 4$) \hspace{.1cm} 3 active + 1 sterile

\hspace{3cm} ($\mu' = 0;\; \mu_3 \neq 0$);
\item ($N_\nu = 5$) \hspace{.1cm} 3 active + 2 sterile

\hspace{3cm} ($\mu' \neq 0;\; \mu_3 = 0$);
\item ($N_\nu = 6$) \hspace{.1cm} 3 active + 3 sterile

\hspace{3cm} ($\mu' \neq 0;\; \mu_3 \neq 0$);
\end{itemize}
In this talk we consider the cases  $N_\nu = 3$ and 4.\cite{brt}

The generalized neutrino mass matrix is then given by~\footnote{This
is similar to the double see-saw mechanism suggested by Mohapatra and
Valle.\cite{mohapatra}}, see figure 9 -
\begin{figure}
\vspace{-1cm}
	\centerline{ \psfig{file=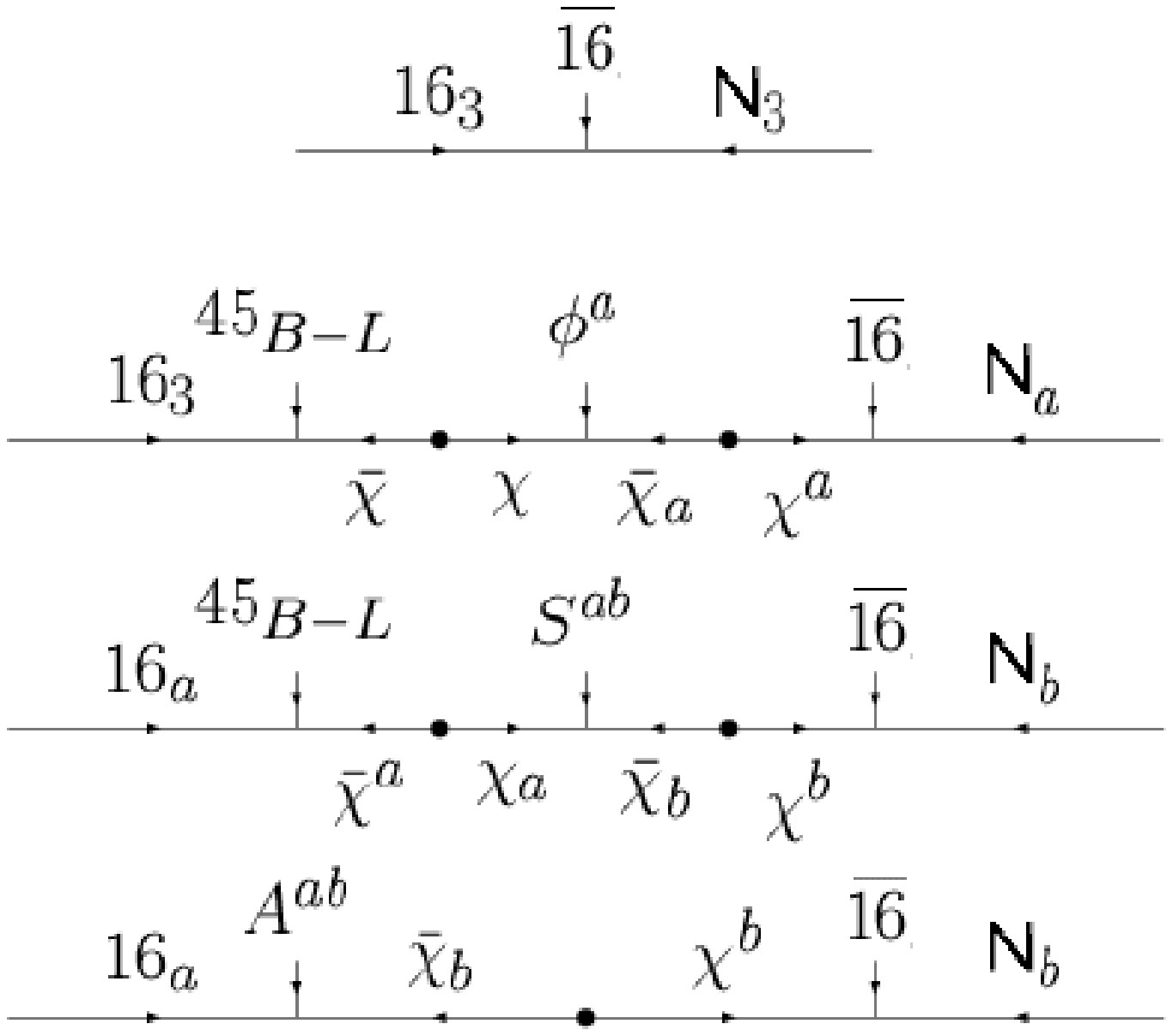,width=10cm,height=21cm}}
\caption{Diagrams generating the neutrino mass matrices}
\end{figure}

\begin{eqnarray}
& ( \begin{array}{cccc}\; \nu & \;\; \bar N & \;\; \bar \nu & \;\;  N
\end{array})  &
\nonumber\\  &  \left( \begin{array}{cccc}  0 & 0 & m & 0  \\
                     0 & 0 & 0 & \tilde \mu^T \\
                     m^T & 0 & 0 & V \\
                     0 & \tilde \mu &\; V^T & M_N  \end{array} \right) &
\end{eqnarray}
where  \begin{eqnarray} m_\nu = & Y_{\nu}\; \langle H_u^0 \rangle
&= \; Y_{\nu}\; {v \over\sqrt{2}}\; \sin\beta \end{eqnarray} and
\begin{eqnarray}
V = & \left( \begin{array}{ccc}  0 & \epsilon' V_{16} & 0 \\
                                - \epsilon' V_{16} & 3 \epsilon V_{16} & 0 \\
                                 0  & r \, \epsilon \, (1 - \sigma) \,
T_{\bar \nu} V_{16}  & \; V'_{16}
\end{array}\right) &
\\
  M_N = & \left( \begin{array}{ccc}  0 & 0 & 0 \\
                                0 & S & \phi \\
                                 0  & \phi  &  0 \end{array}\right) &
\nonumber
\end{eqnarray}
$V_{16},\; V'_{16}$ are proportional to the vev of $\overline{16}$
(with different
implicit Yukawa couplings) and $S, \; \phi$ are up to couplings the vevs of $
S^{22}, \; \phi^2$, respectively.

Since both $V$ and $M_N$ are of order the GUT scale, the states $\bar \nu,
\; N$ may be integrated out of the effective low energy theory.  In this case, the effective neutrino mass matrix is given (at $M_G$) by~\footnote{In fact, at the GUT scale $M_G$ we define an effective dimension 5 supersymmetric neutrino mass operator where the Higgs vev is replaced by the Higgs doublet  H$_u$ coupled to the entire lepton doublet.  This effective operator is then renormalized using one-loop renormalization group equations to $M_Z$.  It is only then that $H_u$ is replaced by its vev.} (the matrix is written in the ($\nu, \ \bar N$) {\em flavor} basis where charged lepton masses are diagonal)
\begin{eqnarray}
 m_\nu^{eff} =  & \hspace{2in}  &
\end{eqnarray}
\begin{eqnarray}
 \tilde U_e^\dagger \; \left( \begin{array}{cc}
m_\nu \;(V^T)^{-1}\;M_N\; V^{-1}\; m_\nu^T & \; - m_\nu \;(V^T)^{-1}\; \tilde \mu\\
                    - {\tilde \mu}^T \; V^{-1}\; m_\nu^T & 0  \end{array}
\right) \; \tilde U_e^* \nonumber
\end{eqnarray}
with
\begin{eqnarray} \tilde U_e = & \left(\begin{array}{cc} U_e & 0 \\
                                               0 & 1 \end{array}\right) & \\
e_0 =  \; e \; U_e^\dagger \;\; ; &  \nu_0 = \; \nu \; U_e^\dagger &  \nonumber
\end{eqnarray}
 $U_e$ is the $3\times3$ unitary matrix for left-handed leptons needed to
diagonalize $Y_e$ (eqn. \ref{eq:yukawa}) and $e_0,\; \nu_0 \; (e, \; \nu)$
represent the
three families of left-handed leptons in the weak- ( mass- )
eigenstate basis for charged leptons.

The neutrino mass matrix is diagonalized by a unitary matrix $U =
U_{\alpha\, i}$;
\begin{eqnarray}
m^{diag}_{\nu} = & U^\dagger \; m^{eff}_{\nu} \; U^* &
\end{eqnarray}
where, in this case, $\alpha= \{\nu_e ,\; \nu_\mu ,\; \nu_\tau ,\; \nu_{s_1}, \;
\nu_{s_2}, \; \nu_{s_3} \}$ is the flavor index and
$i = \{ 1, \cdots, 6\}$ is the neutrino mass eigenstate index.  Recall,
$U_{\alpha\, i}$  is observable in neutrino oscillation experiments.

In general, neutrino masses and mixing angles have many new parameters so
that one might expect to have little predictability.   However, as we shall now see, the U(2)$\times$U(1) family symmetry of the theory provides a powerful
constraint on the form of the neutrino mass matrix.  In particular, the matrix has many zeros and few arbitrary parameters.  Before discussing the four neutrino case, we show why 3 neutrinos cannot work without changing the model.

\subsection{Three neutrinos}

Consider first $m^{eff}_{\nu}$ for three active neutrinos.  We find (at $M_G$) in
the ($\nu_e,\; \nu_\mu,\; \nu_\tau$) basis
\begin{eqnarray}
m^{eff}_{\nu} =  m' \; U_e^\dagger \; \left(\begin{array}{ccc}  0 & 0 & 0 \\
                                       0 & b & 1  \\
                                       0 & 1 & 0  \end{array} \right) \; U_e^*
\end{eqnarray}
with
\begin{eqnarray}  m' &=& \frac{\lambda^2 v^2 \sin^2\beta~ \omega \phi}
{2 V_{16} V'_{16}} \approx {m_t^2 \ \omega \ \phi  \over
V_{16} \ V'_{16}}  \label{eq:3nu} \\
b &= &  \omega \ {S \ V'_{16} \over \phi \ V_{16}} + 2 \ \sigma \ r \ \epsilon
 \nonumber
   \end{eqnarray}
where in the approximation for $m'$ we use
\begin{eqnarray}   m_t (\equiv m_{top}) \approx \lambda \ {v \over \sqrt{2}}
\ \sin\beta ,
\end{eqnarray}  valid at the weak scale.

$m_{\nu}$ is given in terms of two independent parameters \{ $m', \; b$ \}.
Note, this theory in principle solves two problems associated with neutrino
masses.   It naturally has small mixing between $\nu_e - \nu_{\mu}$ since
 the mixing angle comes purely from diagonalizing the charged
lepton mass matrix which, like quarks, has small mixing angles.   While, for
$b \leq  1$, $\nu_{\mu} - \nu_{\tau}$ mixing is large without fine tuning.
Also note, in this theory one neutrino (predominantly $\nu_e$) is
massless.

Unfortunately this theory cannot simultaneously fit both solar and atmospheric
neutrino data.\footnote{We have checked however that we can fit both atmospheric and LSND data.}  This problem can be solved at the expense of
 adding new family symmetry breaking vevs~\footnote{Such an additional vev
 was necessary in the analysis of Carone and Hall.\cite{u2neutrino}}
 \begin{eqnarray}  \langle S^{1 1} \rangle/\kappa_1 & =   \langle S^{1 2} \rangle/\kappa_2 & = \langle S^{2 2}\rangle  .
  \label{eq:kappa} \end{eqnarray}
These are the most general flavor symmetry breaking vevs.  We discuss these three neutrino solutions in a future paper.\cite{brt2}  With $\kappa_{1, 2} \neq 0$ the massless eigenvalue in the neutrino mass matrix is  now lifted.  This allows us to obtain a small mass difference between the first and second mass eigenvalues which was unattainable before in the large mixing limit for $\nu_\mu - \nu_\tau$.  Hence a good fit to both solar and  atmospheric neutrino data can now be found for small values of $\kappa_{1, 2}$.  In fact, only very small values of $\kappa_{1, 2}$ are consistent with charged fermion masses.  We note that with the addition of these two complex parameters the theory is no longer predictive.  In the neutrino sector any of the three solar neutrino solutions (SMA, LMA or "Just so") can be obtained.\cite{brt2}

In the next section we discuss a four neutrino solution to both solar and
atmospheric neutrino oscillations in the theory with $\kappa_{1, 2} = 0$.

 \subsection{Neutrino oscillations   [ 3 active + 1 sterile ]}

 In the four neutrino case the mass matrix (at $M_G$) is given
by~\footnote{This
expression defines the effective dimension 5 neutrino mass operator at
$M_G$ which
is then renormalized to $M_Z$ in order to make contact with data.}

 \begin{eqnarray} m' \  \left[  \begin{array}{cccc}
     U_e^\dagger \,   \left( \begin{array}{ccc}   0 & 0 & 0 \\
                                       0 & b  & 1  \\
                                       0 & 1 & 0  \end{array} \right) \, U_e^* &
- U_e^\dagger \left( \begin{array}{c} 0 \\ u \, c \\ c \end{array} \right) \\
- \left( \begin{array}{ccc} 0 & u \, c & c \end{array} \right) \, U_e^*  & 0
\end{array}
\right]
\end{eqnarray}
where  $m'$ and $b$ are given in eqn. \ref{eq:3nu}  and
\begin{eqnarray}  u &=& \sigma \, r \, \epsilon \label{eq:c} \\
   c &= & {\sqrt{2} \  \mu_3 \, V_{16} \over \omega \,  \lambda \ v \
	\sin\beta \ \phi}
 \approx  { \mu_3 \, V_{16} \over \omega \,  m_t \ \phi}  \nonumber
\end{eqnarray}

In the analysis of neutrino masses and mixing angles we use the fits for
charged fermion
masses as input.  Thus the parameter $u$ is fixed.  We then look for the
best fit to solar
and atmospheric neutrino oscillations.   For this we use the latest
Super-Kamiokande data for atmospheric neutrino oscillations~\cite{atm} and
the best
fits to solar neutrino  data including the possibility of ``just so" vacuum
oscillations or both large and small angle MSW oscillations.\cite{solar}
Our best fit
is found in tables \ref{t:4numass2} and \ref{t:4nuangles}.  It is obtained
in the
following way.

For atmospheric neutrino oscillations we have evaluated the probabilities
($P(\nu_\mu \rightarrow \nu_\mu)$,   $P(\nu_\mu \rightarrow \nu_x) \ {\rm
with} \ x = \{ e, \ \tau, \ s \}$)  as a function of ${\rm x} \equiv
\rm{Log}[(L/km)/(E/GeV)]$.  In order to smooth out the oscillations we have 
averaged the result over a bin size, $\Delta$x = 0.5.   In  fig. 10a we have 
compared the results of our model with a 2 neutrino oscillation model.  We 
see that our result is in good agreement with the
values of $\delta m^2_{atm}$ and $\sin^2 2 \theta_{atm}$ as given.
\renewcommand{\thefigure}{10 \alph{figure}}\setcounter{figure}{0}
\begin{figure}
	\centerline{ \psfig{file=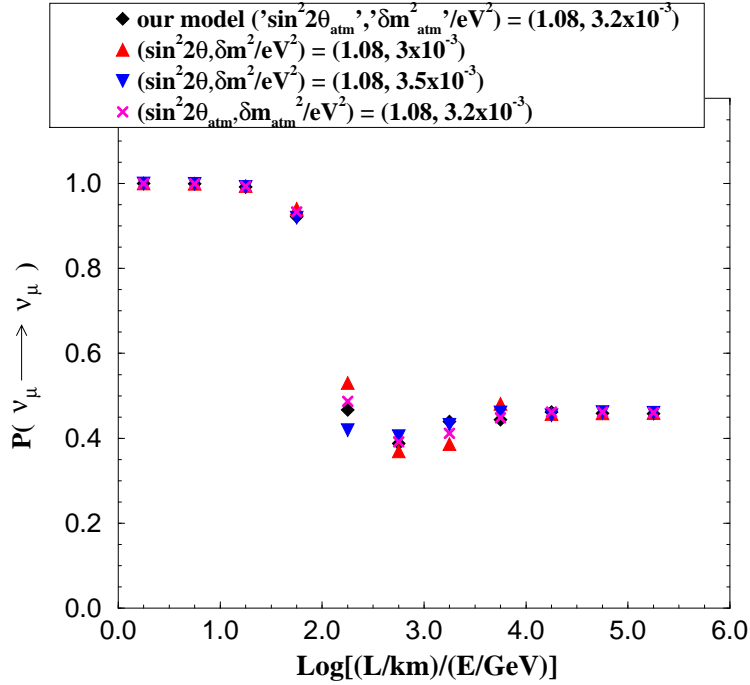,width=9cm,angle=-90}}
\caption{Probability $P(\nu_\mu \longrightarrow \nu_\mu)$ for atmospheric
neutrinos as a function of ${\rm x} \equiv
\rm{Log}[(L/km)/(E/GeV)]$. This probability is averaged in x over bin sizes
 $\Delta {\rm x} = 0.5$.For this analysis, we neglect the matter effects.}
\end{figure}
\begin{figure}
	\centerline{ \psfig{file=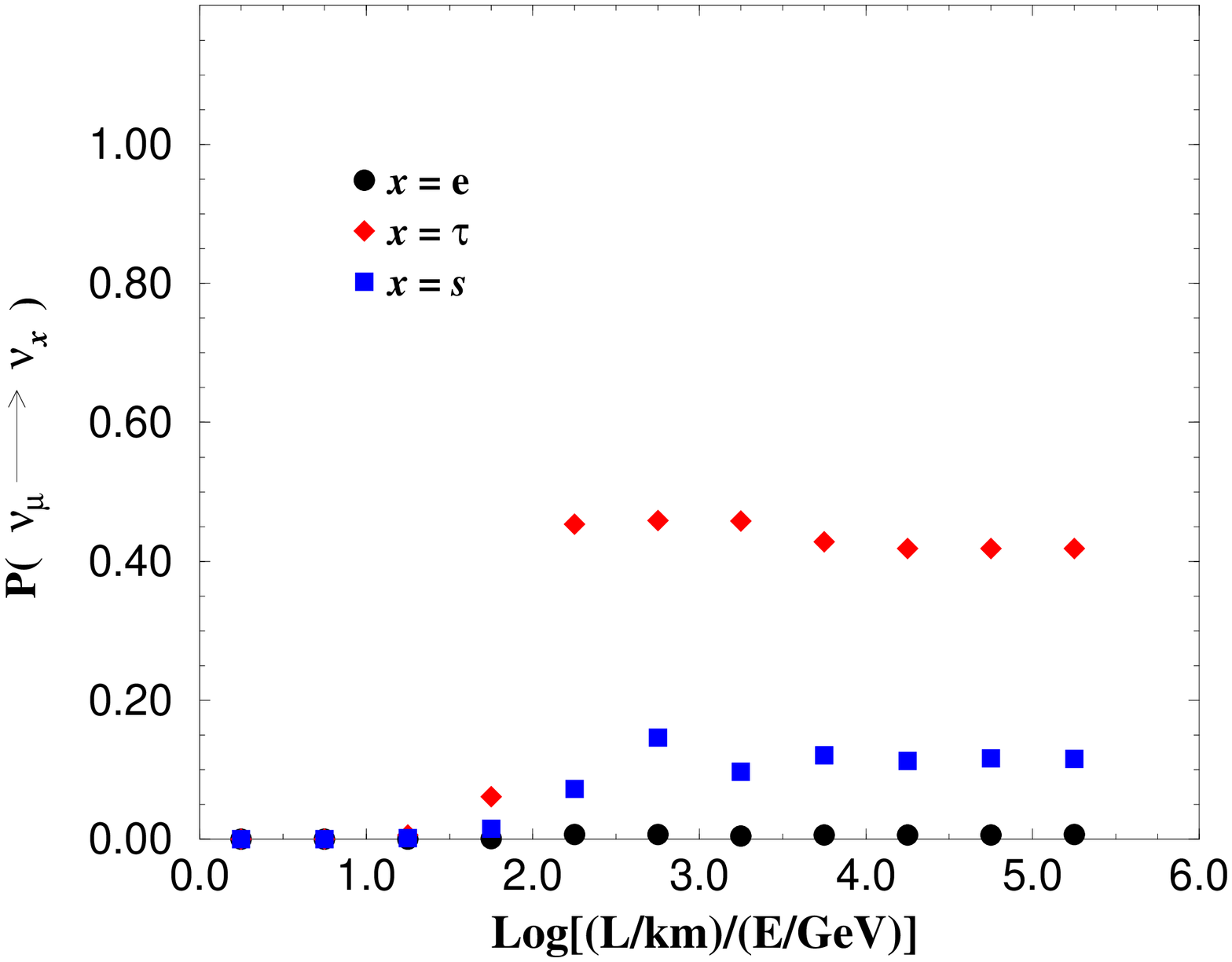,width=9cm,angle=-90}}
\caption{Probabilities $P(\nu_\mu \longrightarrow \nu_x)$ ($x=e$, $\tau$ and
$s$) for atmospheric neutrinos}
\end{figure}

An approximate formula for the effective atmospheric mixing angle is defined by
\begin{eqnarray}
P(\nu_\mu \rightarrow \nu_\mu) \equiv 1- `\sin^2 2 \theta_{atm}` \sin^2
({`\delta m^2_{atm}` \ \rm L \over 4 \ E})  \end{eqnarray}
 with
\begin{eqnarray} `\sin^2 2 \theta_{atm}` &\approx &  4 \ [ \ \| U_{\mu 4}\|^2
( 1 - \| U_{\mu 4} \|^2) \\ 
&+&\| U_{\mu 3} \|^2 ( 1 - \| U_{\mu 3} \|^2 - \| U_{\mu 4} \|^2 )\ ]  \nonumber
\end{eqnarray}
using the approximate relation
\begin{eqnarray}  `\delta m^2_{atm}` = \delta m^2_{43} \approx \delta m^2_{42}
\approx \delta m^2_{41}
\approx \delta m^2_{32} \approx \delta m^2_{31}   .
\end{eqnarray}
Note, $`\sin^2 2 \theta_{atm}` $ may be greater than one.  This is
consistent with the definition above and also with Super-Kamiokande data 
where the best fit occurs for $\sin^2 2\theta_{atm} = 1.05$.  
In a two neutrino oscillation model, values of $sin^2 2\theta > 1$ are unphysical and lead to negative probabilities.  However in our four neutrino model, values of $`\sin^2 2 \theta_{atm}` > 1$ are perfectly physical.    In figure 11, we compare the {\em un-averaged} probability $P(\nu_\mu \rightarrow \nu_\mu)$ for the two neutrino model and the four neutrino model discussed in this talk. 
\renewcommand{\thefigure}{11}
\begin{figure}
	\centerline{ \psfig{file=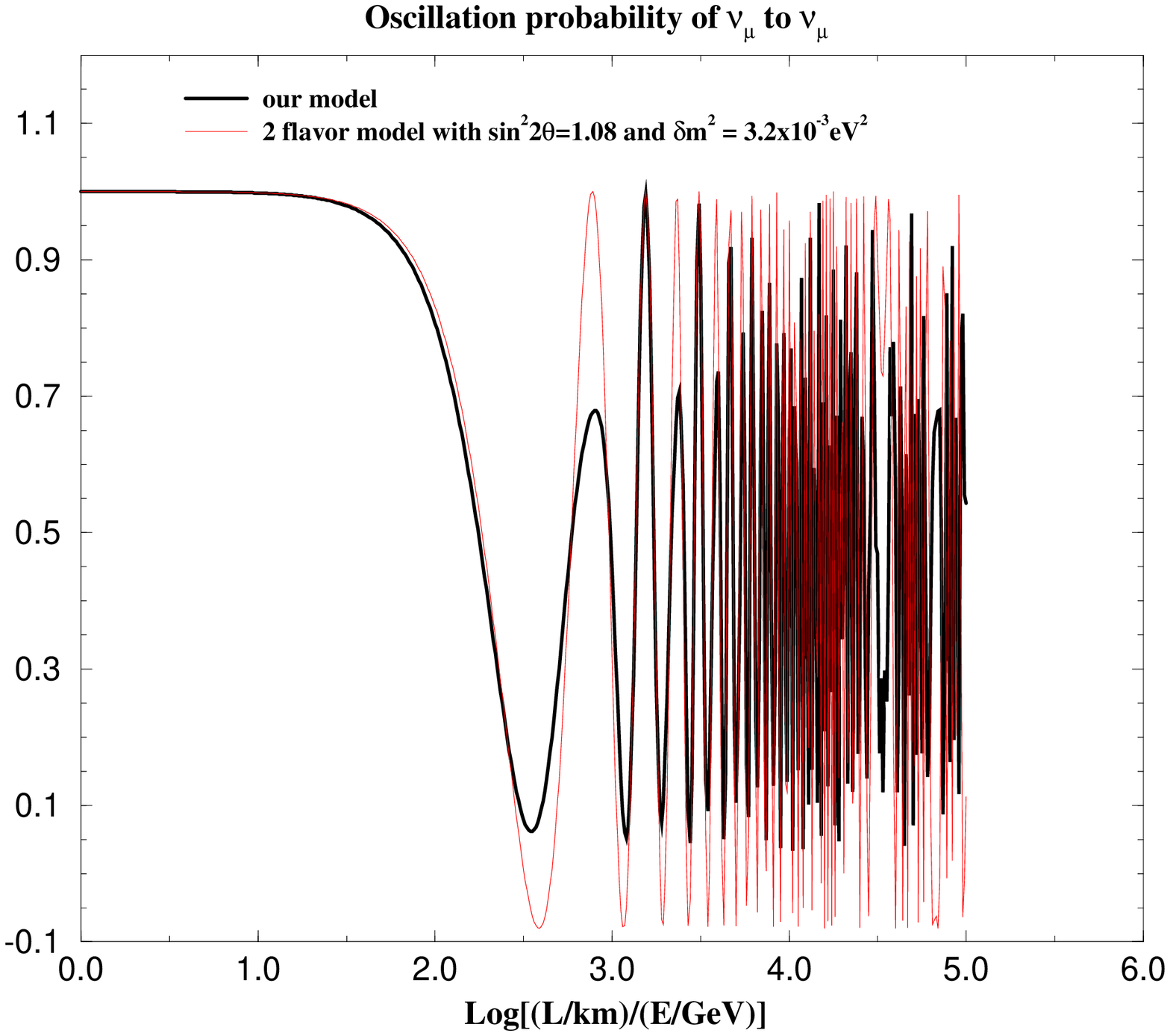,width=9cm, angle=-90}}
\caption{The probability $P(\nu_\mu \rightarrow \nu_\mu)$ for atmospheric neutrinos as a function of ${\rm x} \equiv
\rm{Log}[(L/km)/(E/GeV)]$ without averaging in x.}
\end{figure}

 How can they give equivalent results?  Experimentalists average the oscillating probability $P(\nu_\mu \rightarrow \nu_\mu)$ over bins in the variable ${\rm x} \equiv \rm{Log}[(L/km)/(E/GeV)]$.  This averaging process erases the unphysical negative probabilities of the two neutrino oscillation model.  It is the averaged result which compares well with the physical probability in a model with more neutrinos.  Since it is much simpler to analyze the data in terms of two neutrino oscillations, it is important for experimentalists to allow for unphysical values of $\sin^2 2\theta_{atm}$ in their global fits. These unphysical values may just be an indication of additional neutrinos.

In fig. 10b we see however that although the atmospheric neutrino deficit 
is predominantly due to the maximal mixing
between $\nu_\mu - \nu_\tau$, there is nevertheless a significant ($\sim$ 10\%
effect)  oscillation of $\nu_\mu - \nu_s$.  This effect may be observable at
Super-Kamiokande once the ratio $R_{(NC/CC)}$ (see \ref{eq:R1}, \ref{eq:R2}) is measured more accurately.

The oscillations $\nu_\mu \rightarrow \nu_\tau$ or $\nu_s$ may also be
visible at long baseline neutrino experiments.  For example at K2K~\cite{k2k},
  the mean neutrino energy $E = 1.4 $GeV and distance $L = 250$ km
corresponds to a value of x = 2.3 in fig. 2b and hence  $P(\nu_\mu \rightarrow
\nu_\tau) \sim .4$ and $P(\nu_\mu \rightarrow \nu_s) \sim .1$.  At 
Minos~\cite{minos} low energy beams with hybrid emulsion detectors
 are also being considered. 
These experiments can first test the hypothesis of muon neutrino oscillations
by looking for muon neutrino disappearance (for x = 2.3 we have
$P(\nu_\mu \rightarrow \nu_\mu) \sim .5$).  Verifying oscillations into tau or sterile neutrinos is however much more difficult.
For example at K2K, if only quasi-elastic muon neutrino interactions (single 
ring events at SuperK) are used, then this cannot be tested.  
Minos, on the other hand, may be able to verify the oscillations into tau or sterile neutrinos by using the ratio of neutral current to charged current
measurements~\cite{minos} (the so-called T test). 
 
 \protect
\begin{table}
\caption[3]{
{\bf Fit to atmospheric and solar neutrino \\ oscillations} \\
   \mbox{Initial parameters: ( 4 neutrinos \  $w$/ large tan$\beta$  ) }\ \
\ \

$m' = 7.11 \cdot 10^{-2}$ eV , \ $b$ = $-0.521$, \ $c$ = 0.278,\ $\Phi_b$ =
3.40rad
% end of caption
}
\label{t:4numass2}
$$
\begin{array}{|c|c|}
\hline
{\rm Observable} &{\rm Computed \;\; value} \\
\hline
\delta m^2_{atm}            &  3.2 \cdot 10^{-3} \ \rm eV^2          \\
\sin^2 2\theta_{atm}            &  1.08        \\
 \delta m^2_{sol}   &  4.2\cdot 10^{-6}  \ \rm eV^2  \\
\sin^2 2\theta_{sol} &  3.0\cdot 10^{-3}         \\
\hline
\end{array}$$
\end{table}

\protect
\begin{table}
\caption[3]{
{\bf Neutrino Masses and Mixings} \hspace{1.1cm} \\

\mbox{Mass eigenvalues [eV]: \ \  0.0, \ 0.002, \ 0.04, \ 0.07 \hspace{1cm}} \\
\mbox{Magnitude of neutrino mixing matrix  U$_{\alpha i}$ \hspace{1.7cm}}\\
\mbox{ $i = 1, \cdots, 4$ -- labels mass eigenstates. \hspace{1.5cm}} \\
\mbox{ $\alpha = \{ e, \ \mu, \ \tau, \ s \}$ labels flavor eigenstates.}
% end of caption
}
\label{t:4nuangles}
$$
\left[ \begin{array}{cccc}
0.998                   &  0.0204      & 0.0392   & 0.0529  \\
0.0689                  &    0.291     & 0.567    & 0.767  \\
0.317\cdot 10^{-3}      &  0.145       & 0.771    & 0.620  \\
0.284\cdot 10^{-3}      &   0.946      &  0.287     &  0.154 \\
\end{array} \right]$$
\end{table}

For solar neutrinos we plot,  in figs. 12(a,b), the probabilities  ($P(\nu_e
\rightarrow \nu_e)$,  $P(\nu_e \rightarrow \nu_x) \ {\rm with}
\ x = \{ \mu, \ \tau, \ s \}$) for neutrinos produced at the center of the
sun to propagate
to the surface (and then without change to earth), as a function of the
neutrino energy  E$_\nu$ (MeV).~\footnote{For this calculation
we assume that electron ($n_e$) and neutron ($n_n$) number densities
at a distance $r$ from the center of the sun are given by
$(n_e,~n_n)=(4.6,2.2)\times 10^{11} \exp(-10.5 \frac{r}{R})$ eV$^3$
where $R$ is a solar radius. We also use an analytic approximation necessary to
account for both large and small oscillation scales. For the details, see
the forthcoming paper.\cite{brt}}  We compare our model to a 2 neutrino
oscillation model
with the given parameters.   We see that the solar neutrino deficit is
predominantly due to the small mixing angle MSW solution for $\nu_e -
\nu_s$ oscillations.  The results are summarized in tables
\ref{t:4numass2} and \ref{t:4nuangles}.

A naive definition of the effective  solar mixing angle is given by
\begin{eqnarray} \sin^2 2 \theta_{12} \equiv  4 \ \| U_{e 1} \|^2 \ \|
U_{e 2} \|^2 .
\end{eqnarray}
In fig. 12a we see that the naive definition of $\sin^2 2 \theta_{12} $
underestimates the value of the effective 2 neutrino
mixing angle.   Thus we see that our model reproduces the neutrino results for
$\delta m^2_{sol} = \delta m^2_{12} = 4.2 \times 10^{-6}$ eV$^2$ but instead is
equivalent to a 2 neutrino mixing angle
$\sin^2 2 \theta_{sol} =  3 \times 10^{-3}$
instead of $\sin^2 2 \theta_{12} = 1.6 \times 10^{-3}$.

In addition, whereas the oscillation
$\nu_e - \nu_s$ dominates we see in fig 12b that there is a sigificant
($\sim$ 8\% effect) for  $\nu_e - \nu_\mu$.  This result may be observable at
SNO~\cite{sno}  with threshold $E \ge 5$ MeV for which $P(\nu_e \rightarrow
\nu_\mu) \sim .05$.

\renewcommand{\thefigure}{12 \alph{figure}}\setcounter{figure}{0}
\begin{figure}
	\centerline{ \psfig{file=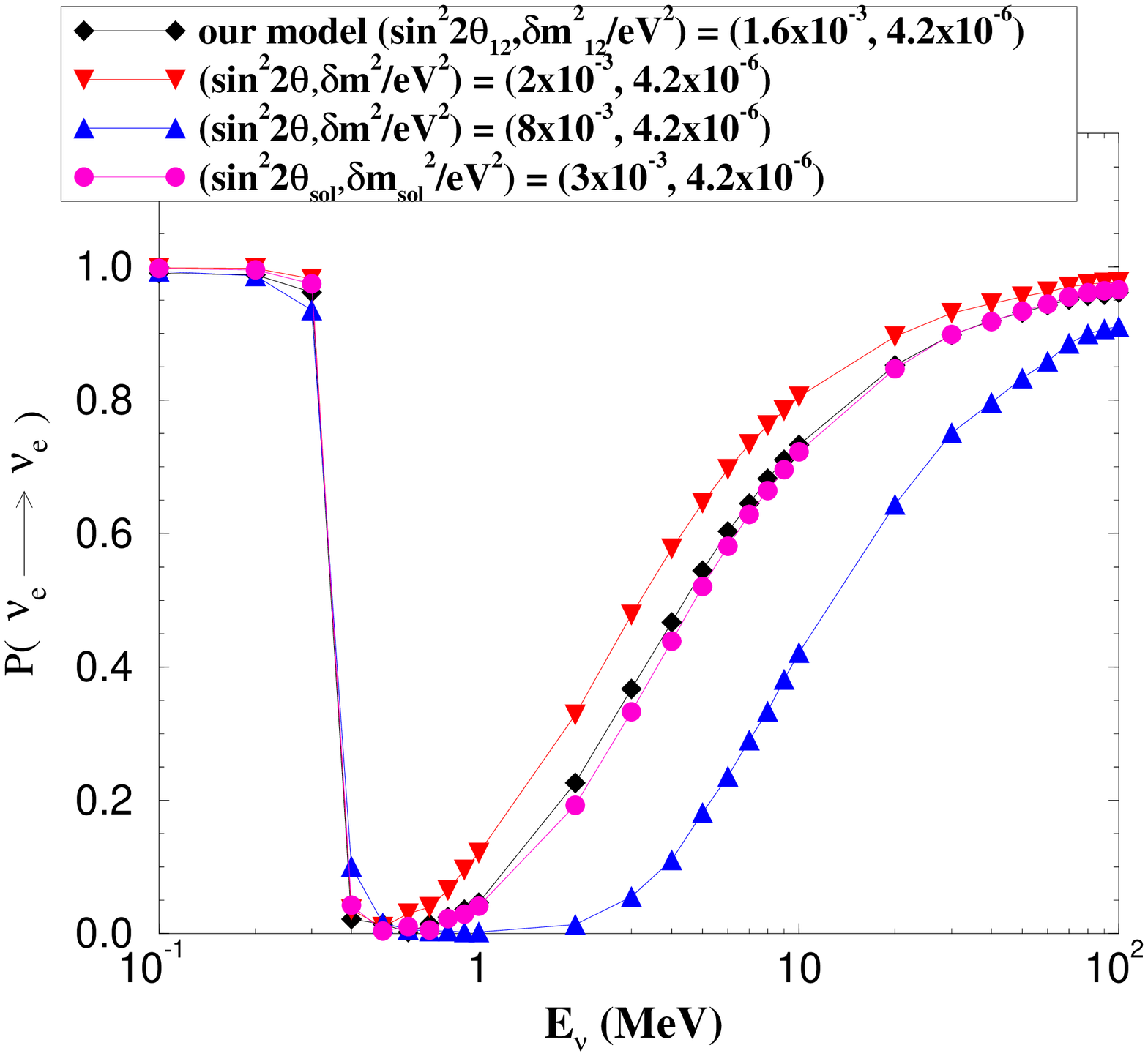,width=9cm,angle=-90}}
\caption{Probability $P(\nu_e \longrightarrow \nu_e)$ for solar neutrinos}
\end{figure}
\begin{figure}
	\centerline{ \psfig{file=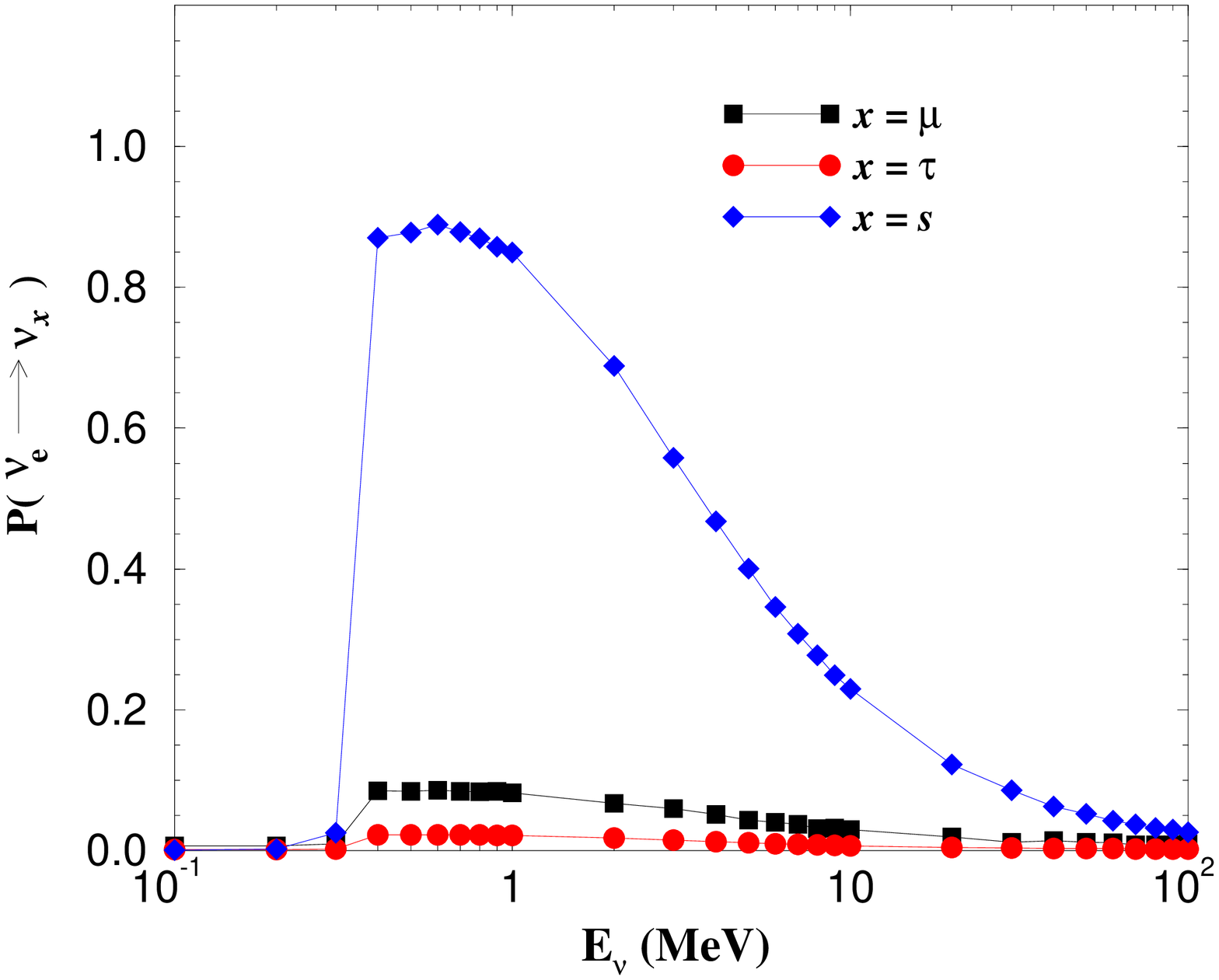,width=9cm,angle=-90}}
\caption{Probabilities $P(\nu_e \longrightarrow \nu_x)$ ($x=\mu$, $\tau$ and
$s$)
for solar neutrinos}
\end{figure}
These results may be compared with a recent analysis by Bahcall, Krastev and Smirnov  [BKS98]\cite{bks}, see figure 13.  The shaded area is the 99\% CL fit for sterile neutrinos consistent with the total rates, zenith-angle
and recoil electron energy spectrum.  The large dot is the best fit and the cross is our solution.
\renewcommand{\thefigure}{13}
\begin{figure}
	\centerline{ \psfig{file=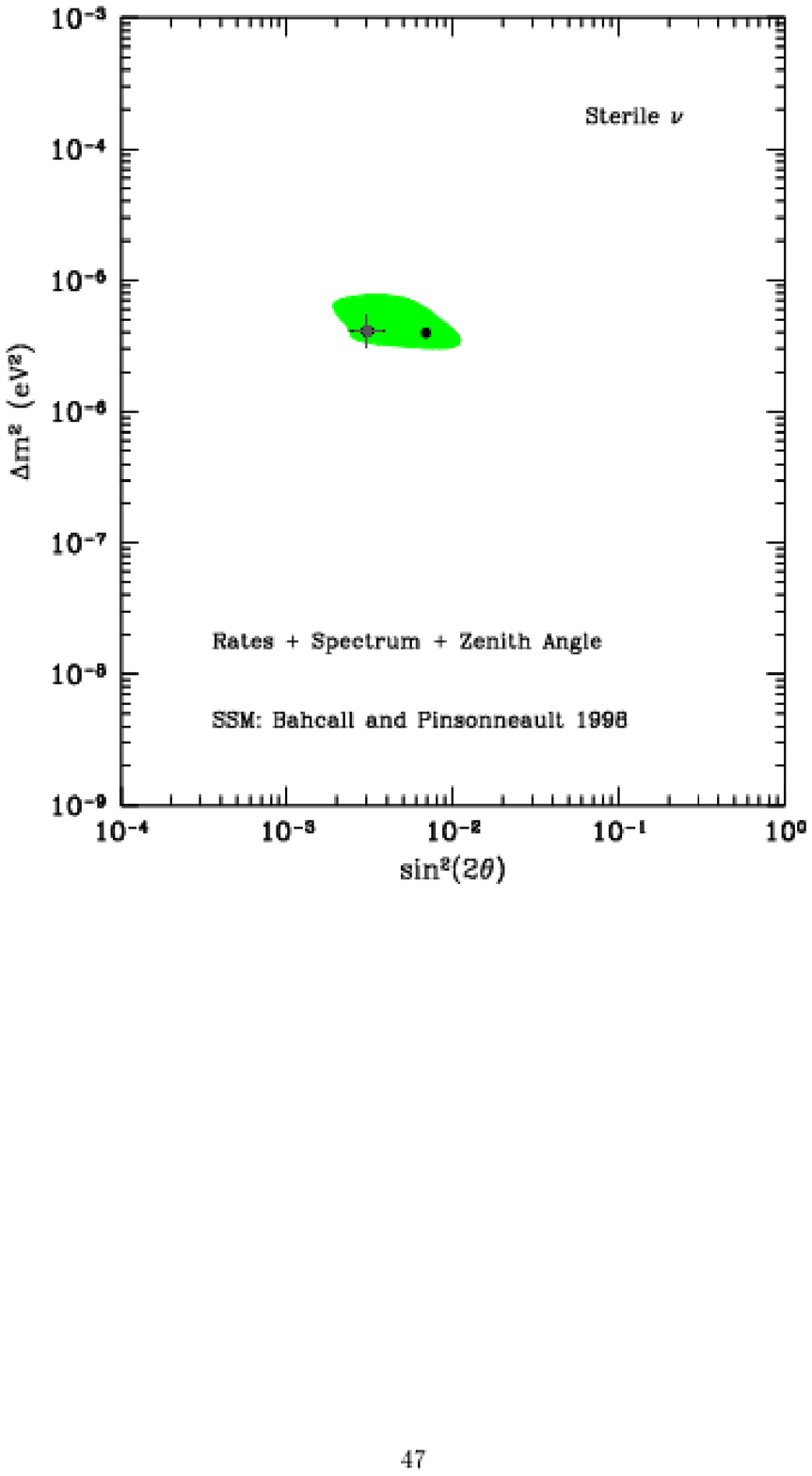,width=9cm,height=21cm}}
\caption{Fit to solar neutrino oscillations by BKS 98.}
\end{figure}

{\em We note that, even though we have four neutrinos, we are {\bf not} able to
simultaneously fit atmospheric, solar and LSND data, i.e. it is not
possible to get both $``\delta m^2_{\nu_e - \nu_\mu}"$ and $sin^2 2\theta$ large enough to be consistent with LSND.}  It is worth asking however how robust is this result.  Not surprisingly we find that upon introducing the new parameters $\kappa_{1, 2}$ (eqn. \ref{eq:kappa}) a solution to all three -- atmospheric, solar and LSND oscillations is obtained.\cite{brt2}  

Finally let's discuss whether the parameters necessary for the fit make sense.
We have three arbitrary parameters.  We have taken $b$ complex,
while any phases for $m'$ and $c$ are unobservable.
A large mixing angle for $\nu_\mu - \nu_\tau$ oscillations is
obtained with $|b| \sim 0.5$.  This does not require any fine tuning; it is
consistent with ${S \ V'_{16} \over \phi \ V_{16}} \sim 1$ which is perfectly
natural (see eqn. \ref{eq:3nu}).
The parameter $c [\rm{eqn.} \ref{eq:c}] \approx 0.28 \approx
{ \mu_3 \, V_{16} \over \omega \,
m_t \ \phi}$ implies $ \mu_3 \approx 0.41 \ {\phi \over V_{16}} \ m_t$.
Thus in order to have a light sterile neutrino we need the parameter $\mu_3
\sim 70$ GeV for $\phi \sim V_{16}$.   Considering that the standard $\mu$
parameter
(see the parameter list in the captions to table \ref{t:fit4nu}) with value
$\mu = 110$ GeV and $\mu_3$ [eqn. \ref{eq:mu'}] may have similar origins,
both generated once SUSY
is spontaneously broken, we feel that it is natural to have a light sterile
neutrino.  Lastly consider the overall scale of
symmetry breaking, i.e. the see-saw scale.  We have $ m' =  7 \times
10^{-2} \rm eV [table ~\ref{t:4numass2}] \approx
{m_t^2 \ \omega \ \phi  \over  V_{16} \ V'_{16}} $.  Thus we find
$ {V_{16} \ V'_{16} \over \phi} \sim  {m_t^2 \ \omega \over m'} \sim
6 \times 10^{14}$ GeV.  This is perfectly reasonable for
$\langle \overline{16} \rangle \sim \langle \phi^2 \rangle \sim M_G$
once the implicit Yukawa couplings are taken into account.

\section{Conclusions and future tests}
We have discussed "predictive" theories of charged fermion masses in this talk.  These theories are constrained by lots of symmetry, in particular SUSY grand unification and family symmetries.  We have argued that such "predictive" theories may constrain neutrino masses and mixing angles as well. And vice versa, neutrino data will therefore be able to constrain theories of charged fermion masses.   

We discussed a particular model with an $SO_{10} \times U_2 \times U_1 \times \cdots$ symmetry.  {\em With minimal family symmetry breaking vevs of the form
$\phi^2 \sim S^{2 2}> A^{1 2}$ this theory is predictive.}  It leads to a very simple picture for neutrino oscillations.  We find maximal $\nu_\mu \rightarrow \nu_\tau$ mixing for atmospheric neutrino oscillations; a small mixing angle MSW solution with $\nu_e \rightarrow \nu_{sterile}$ mixing for solar oscillations and NO solution for LSND.     

We also made the important point that the origin of light sterile neutrinos in SUSY theories may be related to the origin of the Higgs $\mu$ term.  In order to obtain one light sterile neutrino in our model we needed a supersymmetric mass term of the form $\mu_3 N_3 \bar N^3$ where both $N_3,\, \bar N^3$ are $SO_{10}$ singlets and $\mu_3$ is of order $\mu$.  

Our model will be tested in future neutrino experiments.  
\begin{itemize}
\item For atmospheric neutrino oscillations we have $\nu_\mu \rightarrow \nu_\tau$.   Super-Kamiokande will be able to distinguish between $\nu_\tau$ and $\nu_{sterile}$ by looking at the ratio of neutral current to charged current processes.   A recent analysis by Super-Kamiokande\cite{scholberg} measures the ratio $R_{(NC/CC)}   \equiv  \frac{(\pi^0/e)_{Data}}{(\pi^0/e)_{MonteCarlo}}$ consistent with one with large errors.  This result favors $\nu_\tau$ but the result is not yet significant.  In addition, by looking at the zenith-angle dependence one may be able to distinguish between $\nu_\tau, \nu_s$ as well since for $\nu_s$, but not $\nu_\tau$, there is an MSW effect in the earth.
Finally, K2K\cite{k2k} and MINOS\cite{minos} will both be able to confirm the disappearance of $\nu_\mu$ and MINOS should eventually be able to see the appearance of $\nu_\tau$.
\item For solar neutrino oscillations we find a small mixing angle MSW solution with $\nu_e \rightarrow \nu_s$.  Both Super-K and Borexino\cite{borexino} should be able to distinguish between small mixing angle, large mixing angle or vacuum oscillation solutions by their characteristic energy dependence and seasonal variation.\cite{bks}  In addition, SNO should be able to distinguish between active and sterile neutrinos by measuring the ratio of neutral current to charged current solar neutrino processes.   A ratio of order one is indicative of $\nu_e \rightarrow \nu_{active}$, while a ratio much less than one would confirm $\nu_e \rightarrow \nu_{sterile}$ oscillations.  Finally, KAMLAND\cite{kamland} (similar to the CHOOZ experiment) will be sensitive to the large angle MSW oscillation region.  It can thus confirm or rule out this possibility.
\item  We find no evidence for LSND oscillations.   Fortunately, this result will be tested by MiniBOONE\cite{miniboone}; a short base-line oscillation experiment which will be done at Fermilab.
\end{itemize}

\section*{Acknowledgments}
I would like to thank my collaborators T. Blazek and K. Tobe for their help in preparing this talk.  I would also like to thank the organizers of the Conference on Supersymmetry, Supergravity and Strings 99, Seoul National U., Seoul, Korea where this talk was presented and especially to Jihn E. Kim for their kind hospitality. Finally, I would like acknowledge the Institute for Theoretical Physics, Santa Barbara where this talk was written and for partial support from the National Science  Foundation under Grant No. PHY94-07194.    and the DOE under contract DOE/ER/01545-767.

\end{document}